\documentclass[linenumbers]{aastex631}

\usepackage{graphicx,subfigure,color,amsmath,pifont,multirow,threeparttable}
\usepackage{acronym}
\usepackage{mathrsfs}
\usepackage{gensymb}
\acrodef{GRB}{gamma-ray bursts}
\acrodef{GW}{gravitational wave}
\acrodef{EoS}{Equations of State}

\shorttitle{Constraint on EoS of Neutron Star}
\shortauthors{Li et al.}
\begin{document}
	\nolinenumbers
	\title{Revisiting the Constraint on Equation of State of Neutron Star based on the Binary Neutron Star Mergers}
	\correspondingauthor{Da-Bin Lin}
	\email{lindabin@gxu.edu.cn}
	\author{Yun-Peng Li}
	\affiliation{Guangxi Key Laboratory for Relativistic Astrophysics, School of Physical Science and Technology, Guangxi University, \\Nanning 530004, China}
	\author{Zhi-Lin Chen}
	\affiliation{Guangxi Key Laboratory for Relativistic Astrophysics, School of Physical Science and Technology, Guangxi University, \\Nanning 530004, China}
	\author{Da-Bin Lin}
	\affiliation{Guangxi Key Laboratory for Relativistic Astrophysics, School of Physical Science and Technology, Guangxi University, \\Nanning 530004, China}
	\author{En-Wei Liang}
	\affiliation{Guangxi Key Laboratory for Relativistic Astrophysics, School of Physical Science and Technology, Guangxi University, \\Nanning 530004, China}
	\begin{abstract}
		\nolinenumbers
		The merger of neutron star (NS)-NS binary can form different production of the compact remnant, among which the supramassive NS (SMNS) could create an internal plateau and
		the followed steep decay marks the collapse of the SMNS. The proportion of SMNS and the corresponding collapse-time are often used to constrain the NS equation of state (EoS).
		This paper revisits this topic by considering the effect of an accretion disk on the compact remnant, which is not considered in previous works. Compared with previous works,
		the collapse-time distribution (peaks $\sim$100 s) of the SMNSs formed from NS-NS merger
		is almost unaffected by the initial surface magnetic ($B_{{\rm s},i}$) of NS, but the total energy output of the magnetic dipole radiation from the SMNSs depends on $B_{{\rm s},i}$ significantly. Coupling the constraints from the SMNS fraction, we exclude some EoSs and obtain three candidate EoSs, i.e., DD2, ENG, and MPA1. By comparing the distributions of the collapse-time and the luminosity of the internal plateau (in the short gamma-ray bursts) for those from observations with those obtained based on the three candidate EoSs, it is shown that only the EoS of ENG is favored. Our sample based on the ENG EOS and a mass distribution motivated by Galactic systems suggests that approximately $99\%$ of NS-NS mergers collapse to form a black hole within $10^7$s. This includes scenarios forming a
        BH promptly ($36.5\%$), a SMNS ($60.7\%$), or a stable NS that transitions into a BH or a SMNS following
 accretion ($2.1\%$). It also indicates that the remnants for GW170817 and GW~190425, and the second object of GW190814 are more likely to be BHs.
	\end{abstract}

	\section{Introduction}\label{sec:intro}
	The advent of multi-messenger astronomy was marked by the detection of GW~170817, the first gravitational wave (GW) event from a binary neutron star merger, on August 17, 2017 \citep{2017PhRvL.119p1101A}.
	A gamma-ray burst (GRB) GRB~170817A accompanied by a non-thermal afterglow and kilonova (AT~2017gfo) - a thermal emission at optical, near-infrared, and ultraviolet wavelengths - was detected by the $Fermi$ GBM at $\sim1.74$~s post the initial detection \citep{2017ApJ...848L..14G, 2017ApJ...848L..12A, 2017Natur.551...64A, 2017Sci...358.1556C}.
	As progenitors of various astronomical phenomena, such as gravitational waves, gamma-ray bursts, and kilonovae, the NS-NS mergers offer an unprecedented window into these enigmatic occurrences.
	However, the final fate of the merger remnant of GW~170817 remains a subject of ongoing debate, with no consensus achieved thus far.
	Several hypotheses have been proposed to explain the nature of the GW~170817 remnant.
	\cite{2018ApJ...861..114Y} suggested that a long-lived NS with a surface magnetic field of $10^{11}$ to $10^{12}$ G could be the central engine, a hypothesis supported by the observed characteristics of AT~2017gfo.
	This conjecture was backed by \cite{2018ApJ...860...57A}, who, after exploring the permissible parameter space for a long-lasting NS, arrived at a similar conclusion.
	In contrast, \cite{2017ApJ...850L..19M} postulated that the remnant could be a highly massive NS or a short-lived supra-massive NS with a surface magnetic field of $10^{15}$ G, based on their examination of the features of GW~170817.
	\cite{2021ApJ...920..109B} favored a different scenario, proposing a prompt collapse into a black hole.
	These diverse conclusions underscore the ongoing uncertainties surrounding the remnants of NS mergers.
	
	The NS-NS merger is expected to result in various products, depending on the equation of state (EoS) and the total masses of the NS-NS binary. Four potential scenarios have been proposed for the newborn compact object \citep{2010CQGra..27k4105R, 2014MNRAS.439..744R, 2021GReGr..53...59S}:
	(i) the prompt formation of a black hole (BH);
	(ii) the formation of a hypermassive NS collapsing to a BH in $\lesssim$1~s;
	(iii) the formation of a supramassive NS (SMNS) collapsing to a BH in $\sim 10-10^4$~s,
	or (iv) the formation of a stable NS.
	A relativistic jet may also be launched from around the compact remnant
	and be detected as a short GRB (sGRB) if the jet moves along the light of sight.
	The collapse-time and electromagnetic output energy of SMNSs,
	which are expected to collapse due to spin-down, have been extensively studied
	and used to constraint the EoS of NS (\citealp{2014MNRAS.441.2433R, 2014PhRvD..89d7302L, 2016PhRvD..93d4065G, 2021ApJ...920..109B}).
	However, previous studies have neglected the effect of the ejecta fallback on the NS remnants.
	Numerical simulations indicated that the NS-NS merger generates a substantial amount of ejecta, which subsequently falls back onto the surface of the NS.
	The fallback of ejecta has significant effects on the remnant NS,
	affecting its mass, spin, and surface magnetic field (\citealp{2014MNRAS.438..240G, 2021ApJ...907...87L, 1986ApJ...305..235T, 1989Natur.342..656S}).
	Accretion can lead to an increase of the NS's spin and mass, while simultaneously burying the magnetic field lines and consequently reducing magnetic dipole (MD) radiation
	(\citealp{2008Sci...321..376K, 2008MNRAS.388.1729K, 2012ApJ...759...58D, 2021ApJ...917...71Z}).
	Under some certain conditions, however, the accretion onto the NS may be in the propeller regime
	and leads to the spin-down of the NS
	(\citealp{2009ApJ...707.1296P, 2018ApJ...857...95M, 2021MNRAS.502.4680S}).
	Based on a toy model,
	\cite{2022ApJ...939...51M} found that accretion disks indeed impact the angular momentum and mass evolution of neutron stars from a qualitative perspective.
	In this paper, we focus on constraining the NS EoS
	by comparing the NS-NS merger production predicted by EoSs with the sGRB observation, in which the effect of accretion disk is added in our dynamic evolution.

	Our paper is structured as follows.
	In Section~\ref{sec:Method}, we provide a method for sampling the NS-NS systems and distinguishing the associated production, which could be used to preliminarily constrain the equation of state. Furthermore, we present the dynamical evolution of a NS remnant and introduce the sample of sGRB with an internal plateau.
	Section \ref{sec:result} presents the results of the distributions of collapse-times and energy output, which could be used to constrain the EoS (\citealp{2014MNRAS.441.2433R, 2014PhRvD..89d7302L, 2016PhRvD..93d4065G}).
	We discuss the impact of the initial surface magnetic field of the neutron star and the disk mass on the survival time of NS remnants and summarize our findings and conclusions in Section \ref{sec:DISCUSS}.

	\section{Method} \label{sec:Method}
	The NS-NS mergers have four possible scenarios for the newborn compact remnant,
	i.e., (i), (ii), (iii), or (iv) discussed in Section~\ref{sec:intro}.
	During the NS-NS merger, a significant amount of ejecta is expelled and subsequently falls back onto the compact remnant.
	A relativistic jet may also be launched from around the compact remnant
	and be detected as a sGRB if the jet moves along the light of sight.
	In addition, there may be an internal plateau in the light-curve of sGRB's afterglow for the scenario (iii).
	Thus, the sharp decay time of the internal plateau in sGRB
	is nearly equal to the collapse-time of a SMNS.
	The distribution of the collapse-time, associated with the EoSs and NS-NS merger dynamic,
	could be used to constrain the NS EoS.
	In this paper, we revisit this topic by involving the effect of the fallback of the ejecta
	on fate of the compact remnant,
	which is not considered in previous works.
	
	\subsection{NS-NS system sampling and the associated productions} \label{sub1:1}

	\emph{NS-NS system sampling.\;\;}
	In order to obtain the distribution of the collapse-time,
	we extract a sample of $10^5$ NS-NS binary and calculate their post-merger productions.
	The NS-NS binaries are sampled based on the distributions of the chirp mass $M_{\rm{chirp}}$
and the mass ratio ($q$) associated parameter $\tilde{q}=(1-q)/q$ (\citealp{2021ApJ...920..109B}).
Here, $M_{\rm{chirp}}$ follows a normal distribution with mean chirp mass $\overline{{M_{\rm{chirp}}}}=1.175M_{\odot}$
and standard deviation $\sigma_{M_{\rm{chirp}}}=0.044$,
and the parameter $\tilde{q}=(1-q)/q$ follows the exponential distribution $f(\tilde{q})=\lambda_{\rm{\tilde{q}}}e^{-\lambda_{\rm{\tilde{q}}}\tilde{q}}$ with $1/\lambda_{\rm{\tilde{q}}}=0.0954$ (\citealp{2021ApJ...920..109B}).
The above distributions of $M_{\rm{chirp}}$ and $\tilde{q}$ are consistent those of the NS-NS binaries detected in the Milky Way{\footnote {It has not been ruled out that GW~190425 could be the merger of a light NS and a low-mass black hole (\citealp{2020ApJ...891L...5H}). Consequently, GW~190425 is not included in our sampling scheme.}.
Since the vast majority of NS binary systems do not exhibit extreme mass ratios,
we provide a mass ratio $q$ within the range of $0.775-1$.
With the value of both $M_{\rm{chirp}}$ and $\tilde{q}$,
we take $M_1=M_{\rm{chirp}}q^{-3/5}(1+q)^{1/5}$ and $M_2=qM_1$ for the mass of the two NSs in NS-NS binary.	
	\emph{NS-NS merger productions.\;\;}
	Based on $M_1$ and $M_2$, the mass of the dynamical ejecta $M_{\rm{eject,dyn}}$ formed during the NS-NS merger is estimated by
	\begin{equation}
	\frac{M_{\rm{eject,dyn}}}{10^{-3}M_\odot}=(\frac{a_1}{C_{2}}+b_1\frac{M_{1}^{n}}{M_{2}^{n}}+c_1C_{2})M_{2}+(1\leftrightarrow2),
	\end{equation}
	where $C_{2}=GM_2/(R_2c^2)$ is the compactness of $M_2$,
	$G$ is the gravitational constant,
	$c$ is the light speed,
	and $a_1=-9.3335$, $b_1=114.17$, $c_1=-337.56$, and $n=1.5465$ (\citealp{2020PhRvD.101j3002K}).
	
	The total mass of the accretion disk $M_{\rm{disk,total}}$ is estimated by
	\begin{equation}\label{Eq_2}
	M_{\rm{disk,total}}=M_2\times[\max(a_2C_2+c_2,5\times10^{-4})]^{d_2},
	\end{equation}
	where $a_2=-8.1324$, $c_2=1.4820$, and $d_2=1.7784$ (\citealp{2020PhRvD.101j3002K}).
	Equation~(\ref{Eq_2}) is credible when the mass ratio $q$ ranges from 0.775 to 1.
 Seven EoSs covering the mass range of non-rotating neutron star $M_{\rm TOV}$ from low to high are selected. The total baryonic mass is conserved during the NS-NS merger
	and thus the baryonic mass $M_{\rm{s,b}}$ of the compact remnant can be expressed as $M_{\rm{s,b}}=M_{\rm{1,b}}+M_{\rm{2,b}}-M_{\rm{disk,total}}-M_{\rm{eject,dyn}}$,
	where
	$M_{\rm{1,b}}=M_{1}+0.078(R_{1.4M_\odot}/10^5{\rm cm})^{-1}M_{1}^2/{M_\odot}$,
	$M_{\rm{2,b}}=M_{2}+0.078(R_{1.4M_\odot}/10^5{\rm cm})^{-1}M_{2}^2/{M_\odot}$ (\citealp{2020FrPhy..1524603G}), and $R_{1.4M_\odot}$ is the radius of NS with mass 1.4~$M_\odot$ and depends on the EoS as in Table~\ref{tb2:EOS}.
	The baryonic mass $M_{\rm{s,b}}$ can be related to the gravitational mass with $M_{\rm{s,b}}=M_{\rm{s}}+0.078(R_{1.4M_\odot}/10^5{\rm cm})^{-1}M_{\rm{s}}^2/{M_\odot}$.
	The mass of $M_{\rm s}$ is used to distinguish the production of the compact remnant.
	If the compact remnant is a NS, \cite{2018MNRAS.481.3670R} demonstrated through numerical simulations that
	the initial angular momentum should be at the mass shedding limit, i.e.,
	\begin{equation}\label{Eq:Psi}
	\ P_{{\rm max}}=[a_3(\frac{M_{\rm{s,b}}}{M_\odot}-2.5)+b_3]\; \rm ms,
	\end{equation}
	where the values of $a_3$ and $b_3$ depend on the EoS as listed in Table~\ref{tb2:EOS}.
	We assume that NS-NS mergers first result in the formation of a NS at the mass shedding limit, i.e., $P_{{\rm s},i}=P_{{\rm max}}$.
		
	\emph{Fate of the compact remnant.\;\;}
	For a given EoS, the maximum mass of a rotating neutron star with period $P_{\rm{s}}$ can be expressed as \citep{2014PhRvD..89d7302L, 2014MNRAS.441.2433R}
	\begin{equation}
	\ M_{\rm{max}}(P_{\rm{s}})=M_{\rm{TOV}}(1+\alpha P_{\rm{s}}^\beta),
	\label{equation 4}
	\end{equation}
	where the $M_{\rm{TOV}}$ is the maximum NS mass for a nonrotating NS, and the values of $\alpha$ and $\beta$ are listed in Table \ref{tb2:EOS}.
	If the mass of the NS is equal to $M_{\rm{max}}$, the NS will collapse into a BH.
	So we can define a critical period
	\begin{equation}
	P_{\rm{c}}=(\frac{M_{\rm{s}}-M_{\rm{TOV}}}{\alpha M_{\rm{TOV}}})^{1/\beta}.
	\end{equation}
	Based on the value of $M_{\rm s}$, $M_{\rm{max}}(P_{{\rm s},i})$, $M_{\rm{TOV}}$, and $M_{\rm{disk}}$, we can have the following cases:
	\begin{itemize}
		\item \textbf{Case I}.\,
		If $M_{\rm{s}}>M_{\rm{max}}(P_{\rm s,i})$ is satisfied for a newborn compact object,
		the compact remnant would promptly collapse into a BH.
		\item \textbf{Case II}.\,
		If $M_{\rm{s}}+M_{\rm{disk}}<M_{\rm{TOV}}$, the compact remnant would be a stable NS.
		Here, $M_{\rm{disk}} \approx (1-\xi_{\rm{w}})M_{\rm{disk,total}}$
		is the total accretion mass onto the compact object.
		The wind ejecta, which is the outflow of the accretion disk,
		can be estimated as a constant fraction of the disk mass, i.e., $M_{\rm{eject,w}} \approx \xi_{\rm{w}}M_{\rm{disk,total}}$ with $\xi_{\rm{w}} = 0.2$ adopted in this paper (\citealp{2015MNRAS.446..750F}, \citealp{2015MNRAS.448..541J}, \citealp{2017PhRvD..95j1303S}).
		\item \textbf{Case III}.\,
		If $M_{\rm{s}}>M_{\rm{TOV}}$ and $P_{{\rm s,i}}<P_{\rm{c}}$, the compact remnant would be initially a SMNS,
		which may collapse into a BH during the following fallback of the ejecta or spin-down of the SMNS.
		
		\item \textbf{Case IV}.\,
		If $M_{\rm{s}}+M_{\rm{disk}}>M_{\rm{TOV}}$ but $M_{\rm{s}}<M_{\rm{TOV}}$, the compact object would be initially a stable NS. During the following fallback of the ejecta, the stable NS would transform into a SMNS or BH.
	\end{itemize}
	The final fate of the NS remnant in both Case III and Case IV (Cases III-IV)
	is our focus
	and associated with the co-evolution of the NS and the accretion disk,
	which is presented in Section~\ref{sub1:3}.
	
	\begin{table}
		\centering{
			\begin{tabular}{ccccccc}
				\hline \hline
				EoS & $M_{\rm{TOV}}$  & $R_{1.4M_{\odot}}/10^5{\rm cm}$  &  $\alpha/(10^{-10}s^{-\beta}) $ & $\beta$ & $a_3$ & $b_3$
				\tabularnewline
				\hline
				SLy & 2.05$M_{\odot}$ & 11.74 & 1.60 & $-2.75$ & $-0.25$ & 0.72 \\
				APR  & 2.20$M_{\odot}$  & 11.37  & 0.303 & $-2.95$ & $-0.21$ & 0.69 \\
				ENG & 2.24$M_{\odot}$ & 12.06 & 3.95 & $-2.68$ & $-0.20$ & 0.77 \\
				DD2  &  2.42$M_{\odot}$ & 13.12& 1.37 & $-2.88$ & $-0.20$ & 0.93 \\
				MPA1 & 2.46$M_{\odot}$ & 12.47 & 0.53 & $-2.98$ & $-0.17$ & 0.84 \\
				MS1  & 2.77$M_{\odot}$ & 14.70 & 5.88 & $-2.72$ & $-0.21$ & 1.10 \\
				MS1b & 2.78$M_{\odot}$ & 14.46 & 4.09 & $-2.77$ & $-0.20$ & 1.07 \\
				\hline
		\end{tabular}}
		\caption{Parameters of various NS EoS models, which are taken from \cite{2014PhRvD..89d7302L} , \cite{2016PhRvD..94h3010L} , \cite{2018ApJ...860...57A} , \cite{2020A&A...641A..56R} , \cite{2020FrPhy..1524603G} , \cite{2018MNRAS.481.3670R} , and \cite{2009PhRvD..79l4032R}.}\label{tb2:EOS}
	\end{table}

	\subsection{Dynamical Evolution of a remnant Neutron Star and Its Collapse} \label{sub1:3}
	\emph{Alfv$\acute{e}$n radius and corotation radius of NS-Disk system.\;\,}
	The co-evolution of a NS and the accompanied accretion disk is related to the magnetic field of the NS\footnote{\cite{2022ApJ...939...51M} presents a comparative analysis of the timescales
		for the dynamic evolution of the NS remnant
		associated with
		the neutrino viscosity,
		the magnetorotational instability,
		the convection,
		and the Spruit-Tayler dynamo.
		It is suggested that the residual core of the NS remnant entering the solid-body rotation phase
		is earlier than the transfer of angular momentum between the solid-body rotating NS and the disk.
		Then, a solid-body rotating NS remnant accompanied with a a disk is used in this paper.}
	The accretion disk may be in an accretion state or a propeller state,
	depending on the relation of the Alfv$\acute{e}$n radius $R_{\rm{m}}$ and the corotation radius $R_{\rm{c}}$.
	The Alfv$\acute{e}$n radius is the radius at where the pressure from the magnetic field of the NS on the accretion flow and the ram pressure of
	the accretion flow are in balance, i.e.,
	\begin{equation}
	R_{\rm{m}}=\mu^{4/7}(GM_{\rm{s}})^{-1/7}\dot{M}_{\rm{sup}}^{-2/7},
	\end{equation}
	where $\mu=B_{\rm{s}}R_{\rm{s}}^{\rm{3}}$, $B_{\rm{s}}$, and $R_{\rm{s}}$ are respectively the magnetic moment,
	the strength of surface dipolar magnetic field,
	and the radius of the NS (here $R_{\rm{s}}=11.89$~km for DD2, $R_{\rm{s}}=10.77$~km for ENG, $R_{\rm{s}}=11.42$~km for MPA1 are adopted based on RNS code (\citealp{2018ApJ...860...57A,2020A&A...641A..56R}))
	and $\dot{M}_{\rm{sup}}$ is the mass supply rate of accretion flow in the disk.
	If $R_{\rm{m}}$ is larger than $R_{\rm{s}}$, the accretion disk would be truncated at $R_{\rm{m}}$.
	The corotation radius is defined as
	\begin{equation}
	R_{\rm{c}}=(GM_{\rm{s}}/\Omega_{\rm{s}}^2)^{1/3},
	\end{equation}
	where $\Omega_{\rm{s}}={2\rm{\pi}}/P_{\rm{s}}$ is the initial angular frequency of the NS.
	
	\emph{Accretion rate and Torque of NS-Disk system.\;\,}
	Depending on the relation of $R_{\rm{m}}$ and $R_{\rm{c}}$, there are two possible scenarios for the co-evolution of the accretion disk and NS.
	(1)
	In the situation of $R_{\rm{m}}>R_{\rm{c}}$,
	the magnetic field would accelerate the disk material beyond the corotation radius
	and thus produces a propeller outflow $\dot{M}_{\rm pro}$.
	Correspondingly, the NS may spin down.
	(2)
	In the situation of $R_{\rm{s}}<R_{\rm{m}}<R_{\rm{c}}$, the magnetic field of the NS decelerates the disk material within the corotation radius.
	The disk-NS interaction leads to the transfer of angular momentum from the disk to the NS,
	causing the latter to spin up.
	If the accretion rate of the disk onto the NS is $\dot{M}_{\rm acc}$ in these two scenarios,
	the torque acting on the NS due to the accretion can be written as $T_{\rm{acc}}=\dot{M}_{\rm acc}R_{\rm{m}}^2\Omega_{\rm{K}}(R_{\rm{m}})$,
	where $\Omega_{\rm{K}}(R)$ is the Kepler angular velocity of the accretion disk at the radius $R$.	
	In addition, the torque acting on the NS caused by propeller outflow can be written as $T_{\rm{pro}}=-\dot{M}_{\rm pro}R_{\rm{m}}^2\Omega_{\rm{s}}$.
	Following \cite{2021ApJ...907...87L},
	the parameter $\omega=\Omega_{\rm{s}}/\Omega_{\rm{K}}(R_{\rm{m}})=(R_{\rm{m}}/R_{\rm{c}})^{3/2}$ is introduced to unify the description of the above two scenarios,
	i.e.,
	\begin{equation}\label{Eq:M_sup}
	\dot{M}_{\rm{sup}}=\dot{M}_{\rm{acc}}+\dot{M}_{\rm{pro}},
	\end{equation}
	\begin{equation}\label{Eq:M_disk}
	T_{\rm{disk}}=T_{\rm{acc}}+T_{\rm{pro}}=\frac{1-\omega^{n+1}}{1+\omega^n}\dot{M}_{\rm{sup}}\sqrt{GM_{\rm{s}}R_{\rm{m}}}.
	\end{equation}
	where $\dot{M}_{\rm{acc}}=\dot{M}_{\rm{sup}}/(1+\omega^n)$,
	and $\dot{M}_{\rm{pro}}=\omega^n\dot{M}_{\rm{sup}}/(1+\omega^n)$.
	Here, the artificial parameter $n > 1$ is introduced to represent the
	sharpness of the transition between scenarios (1) and (2),
	and $n = 2$ is took in our paper (\citealp{2021ApJ...907...87L}).
	Equation~(\ref{Eq:M_sup}) is reduced to $\dot{M}_{\rm{acc}}\approx\dot{M}_{\rm{sup}}$ and $\dot{M}_{\rm{pro}}\approx0$ for $\omega\ll1$, i.e., scenario~(2);
	and $\dot{M}_{\rm{acc}}\approx0$ and $\dot{M}_{\rm{pro}}\approx\dot{M}_{\rm{sup}}$ for $\omega\gg1$, i.e., scenario~(1).

	\emph{Dynamical Evolution of NS-Disk system.\;\,}
	The dynamical evolution of the newborn NS and the disk can be described as (e.g., \citealp{2014MNRAS.438..240G,2021ApJ...907...87L})
	\begin{equation}\label{EQ_Evo1}
	\frac{dJ_{\rm{s}}}{dt}=T_{\rm{md}}+T_{\rm{gw}}+T_{\rm{disk}},
	\end{equation}
	\begin{equation}
	\frac{dJ_{\rm{disk}}}{dt}=-T_{\rm{disk}}+T_{\rm{rec}},
	\end{equation}
	\begin{equation}
	\frac{dM_{\rm{s,b}}}{dt}=\dot{M}_{\rm{acc}}(t),
	\end{equation}
	\begin{equation}\label{EQ_Evo4}
	\frac{dM_{\rm{disk}}}{dt}=-\dot{M}_{\rm{sup}}+\dot{M}_{\rm{rec}},
	\end{equation}
	where $M_{\rm{s,b}}$ ($M_{\rm{disk}}$) and $J_{\rm{s}}$ ($J_{\rm{disk}}$) are respectively
	the baryonic mass and the angular momentum of the NS (the accretion disk),
	and $\dot{M}_{\rm{rec}}$ is the fallback rate of the propeller outflow, among which $J_{\rm{s}}=I_{\rm{s}}\Omega_{\rm{s}}$.
	Here, $T_{\rm{md}}=-{\mu^2\Omega_{\rm{s}}^3}/{c^3}(1+\rm{sin}^2\chi)$ (\citealp{2006ApJ...648L..51S}) and $T_{\rm{gw}}=-
			(2G I_{\rm{s}}^2\varepsilon^2/5c^5)\Omega_{\rm{s}}^5\rm{sin}^2\chi(1+15\rm{sin}^2\chi)$ (\citealp{2000PhRvD..63b4002C,2015ApJ...798...25D})
			are respectively the torques due to MD radiation and GW radiation,
			where $\chi$ is the angle between the magnetic-field and spin-axes and $\chi=\pi/2$ is adopted,
			$I_{\rm{s}}=0.237M_{\rm{s}}R_{\rm{s}}^2[1+4.2(M_{\rm{s}}/M_{\rm{\odot}})/(R_{\rm{s}}/\rm km)+90(M_{\rm{s}}/M_{\rm{\odot}})^4/(R_{\rm{s}}/\rm km)^4]$ is the inertial moment of the NS (\citealp{2005ApJ...629..979L}),
			and the ellipticity $\varepsilon$ of the NS due to the deformation by the internal magnetic field $B_{\rm int}$
			can be described as $\varepsilon=10^{-4}(B_{\rm int}/10^{16}\rm G)^2$
			and $\varepsilon=0.001$ is typically adopted (\citealp{1992Natur.357..472U}; \citealp{2013PhRvD..88f7304F}).
	To assess whether the dynamical bar-mode instability and secular instability could arise during the evolution of the NS and lead to additional GW radiation, we also calculate the proportion of the NS’s rotational kinetic energy relative to its absolute gravitational binding energy, $\xi=T/|W|$, where $|W|$ can be expressed as (\citealp{2001ApJ...550..426L})
			\begin{equation}
			|W|=\frac{3GM_{\rm{s}}^2}{5R(1-\frac{GM_{\rm{s}}}{2R_{\rm s}})}.
			\end{equation}
			The dynamical bar-mode instability will occur for $\xi>0.27$ (\citealp{1995ApJ...442..259L}), and the secular instability occur for $\xi>0.14$ (\citealp{1995ApJ...442..259L}).
   
During the evolutionary process, if the spin of a neutron star exceeds the mass shedding limit, then the neutron star’s spin will evolve while maintaining the mass shedding limit (\citealp{2022ApJ...939...51M}). The period and spin frequency of NS can be expressed as
			\begin{equation}
			P_{\mathrm{s}} = \left\{
			\begin{array}{ll}
			[a_3\left(\frac{M_{\mathrm{s,b}}}{M_\odot}-2.5\right)+b_3]\times10^{-3} & \text{if } P_{\mathrm{s}} \geq P_{\mathrm{max}} \\
			\frac{2\pi I_{\mathrm{s}}}{J_{\mathrm{s}}}& \text{if } P_{\mathrm{s}} < P_{\mathrm{max}}
			\end{array}
			\right.
			\end{equation}
			and $\Omega_{\rm s}=2\pi/P_{\rm s}$.

	\emph{Initial settings for the NS-disk system.\;\,}
	Following the work of \cite{2008Sci...321..376K}, \cite{2008MNRAS.385.1455M}, and \cite{2017ApJ...849...47L}, we assume that the disk is a torus-like structure located at the radius $R_{\rm torus}$.
	Then, the total mass supplied by the disk (without the effect of the magnetic field of the NS)
	can be estimated with $\dot{M}_{\rm{sup}}=M_{\rm{disk}}/\tau_{\rm{vis}}$,
	where the viscosity timescale $\tau_{\rm{vis}}$ is given as $\tau_{\rm{vis}}\approx 2/[\alpha\Omega_{\rm{K}}(R_{\rm{torus}})]$ and $\alpha$ is the dimensionless viscosity parameter (\citealp{1973A&A....24..337S}).
	Due to gravitational confinement and external material obstruction, the outflow driven by the propeller process
			is assumed to entirely fallback to the disk and thus $\dot{M}_{\rm{rec}}=\dot{M}_{\rm{pro}}$ and $T_{\rm{rec}}=-T_{\rm{pro}}$ are adopted (\citealp{2021ApJ...907...87L}).
	The angular momentum of the disk at time $t$ can be described as $J_{\rm{disk}}(t)\simeq\sqrt{GM_{\rm{s}}R_{\rm{torus}}}M_{\rm{disk}}$ (\citealp{2021ApJ...907...87L}).
	In our calculations, we assume $J_{\rm{disk}}(t=0) = J_{\rm{s}}(t=0)$ and $\alpha=0.05$ is adopted.
	
	\emph{Magnetic field evolution of the NS.\;\,}
	The accretion onto the NS not only affects the NS's mass but also its surface magnetic field.
	The surface dipole magnetic field decreases due to the accretion
	and its evolution is generally described with the following empirical equation (\citealp{1986ApJ...305..235T, 1989Natur.342..656S}):
	\begin{equation}
	B_{\rm{s}}(t)=\frac{B_{{\rm s},i}}{1+M_{\rm{acc}}(t)/M_{\rm{c}}},
	\end{equation}
	where $B_{{\rm s},i}=B_{\rm s}(t=0)$ is the initial strength of the surface dipolar field for the newborn NS, $M_{\rm{acc}}(t)=M_{\rm{s,b}}(t)-M_{\rm{s,b}}(t=0)$ is the total accreted mass of the NS, and $M_c=0.001M_{\odot}$ is adopted in our calculations (e.g., \citealp{2021ApJ...907...87L}).

	\emph{Collapse-time of a NS Remnants and Corresponding Energy output.\;\,}
	Equation~(\ref{equation 4}) reveals that if the mass of the NS equals to or is larger than $M_{\rm{max}}$, the NS would collapse into a BH.
	Correspondingly, one can obtain a collapse-time $t_{\rm col}$ for a NS
	by solving the time of $M_{\rm{s}}=M_{\rm{max}}$
	based on Equations~(\ref{EQ_Evo1})-(\ref{EQ_Evo4}).
	Then,
	the total output energy of the magnetic dipole (MD) radiation is
	(e.g., \citealp{2001ASPC..248..469R})
	\begin{equation}
	E_{\rm{md}}=\int_{0}^{t_{\rm col}} {L_{\rm{md}}} dt=\int_{0}^{t_{\rm col}} {\frac{\mu^2\Omega_{\rm{s}}^4}{6c^3}} dt,
	\end{equation}
	where $L_{\rm{md}}(t)={\mu^2[\Omega_{\rm{s}}(t)]^4}/({6c^3})$.
	The XRT luminosity at the collapse-time $t_{\rm col}$
	can be described as
	\begin{equation}
	L_{\rm{xrt}}=\frac{\eta_{\rm{jet}}\eta_{\rm{xrt}}L_{\rm{md}}(t_{\rm col})}{2(1-\rm{cos\,\theta_{\rm jet})}},
	\end{equation}
	where $\eta_{\rm{jet}}$ represents the efficiency of MD radiation energy injected into the jet,
	$\eta_{\rm{xrt}}$ signifies the transform efficiency of the jet luminosity into the  luminosity, and $\theta_{\rm jet}$ denotes the jet opening angle. 
			In our calculations, we simply assume that only half of the MD radiation energy is injected into the jet. Specifically, we use the formula $\log{\eta_{\rm{xrt}}}=-0.042[\log{L_{\rm{md}}(t_{\rm col})}]^2+3.81\log{L_{\rm{md}}(t_{\rm col})}-87.29$ (\citealp{2019ApJ...878...62X}). We set the jet inject efficiency, $\eta_{\rm{jet}}$, to 0.5, and adopt an typical jet opening angle, $\theta_{\rm{jet}}=5^\circ$ (\citealp{2019MNRAS.483..840B,2020PhR...886....1N}). While a larger jet opening angle would result in lower XRT luminosity, a higher injection efficiency would lead to a higher XRT luminosity.

	\subsection{Sample of sGRBs with an internal plateau}
	There are two types of plateaus observed in sGRBs.
	The first type is the external plateau, characterized by light curves that transition smoothly from
	a flat, slowly decaying phase to a normal decay phase of the external-forward shock.
	More quantitatively,
	the the decline timescale $\Delta  t$ is approximately equal
	to $t$ (where $t$ is the time since the GRB trigger of the end of the plateau), i.e., $\Delta {t}/t\sim1$.
	External plateaus can be explained by emission from the external shock.
	The second type is the internal plateaus,
	which exhibits a rapid decay at the end of the plateau phase,
	i.e., $\Delta {t}/t\ll 1$,
	which is very different from that of the external-forward shock (\citealp{2013MNRAS.430.1061R,2015ApJ...805...89L,2020ApJ...895...58G,2024ApJ...960...17L}).
	It reveals that the central engine drives the internal plateaus
	and the abrupt cutoff is attributed to the collapse of a NS in Case III-IV.
	\cite{2013MNRAS.430.1061R}, \cite{2015ApJ...805...89L} and \cite{2020ApJ...895...58G} analyzed the sGRBs exhibiting internal plateau features, and their results indicate that the sharp decline following plateau phase in these sGRBs is predominantly concentrated within the time range of $100-300$ seconds.
	
	In order to get the observational collapse-time distribution,
	we collected sGRBs with internal plateau from the \cite{2015ApJ...805...89L} and \cite{2020ApJ...895...58G} to obtain the collapse-times $t_{\rm col}$ and then calculated the corresponding luminosities $L_{\rm col}=4\pi D_{l}^2F_{\rm v,col}$ at the moment of collapse, where $D_{l}$ is the luminousty distance.
	For the sGRBs without redshift, we assume a value of $z$=0.54 (the average redshift value of sGRBs with known redshift).
	We then compare the observed collapse-time and luminosity distribution with that of our sample (Section~\ref{sub1:1}), the results presented in Section~\ref{sec:result}.

\section{result} \label{sec:result}
	sGRBs with internal plateau may originate from Cases~III-IV mentioned in Section~\ref{sub1:1}.
	Since sGRBs may be from the merger of BH-NS and the internal plateau may be missed if the external shock emission is dominant in the afterglow,
	the intrinsic proportion of Cases III-IV should be higher than the observed proportion of sGRBs with an internal plateau\footnote{ sGRB may formed from the NS-NS merger, the BH-NS merger, or the other channels (e.g. the collapse of a massive stellar (\citealp{2021NatAs...5..917A,2022ApJ...932....1R})).
We take the contribution fraction of the NS-NS merger to form sGRB as $f_{\rm NS-NS}$ ($<1$).
For the NS-NS mergers, we assume the fraction of Cases~III-IV among Cases~I-IV
as $f_{\rm III-IV}$, which can be estimated based on a specified EoS. 
Since the sGRB with an internal plateau originates from Cases~III-IV,
the fraction of the sGRB with an internal plateau among observed sGRBs
can be estimated as $< f_{\rm III-IV}\times f_{\rm NS-NS}$,
which should be less than $f_{\rm III-IV}$.}.
	Same as \cite{2016PhRvD..93d4065G},
	we employ the value of $22\%$ as the lower bound for the formation fraction of Cases III-IV.
	In Figure~\ref{fig1:EOS}, we plot the formation fraction of Cases III-IV with different EoSs.
	One can find that
	the corresponding formation fraction of Cases~III-IV for the EoSs of DD2, ENG, and MPA1 is consistent with the observations.

	\emph{Collapse-time distribution.\;\,}
	Then, we explore the influence of accretion disks on the NS remnant under these three kinds of EoS.
	In Figure~\ref{fig2:time},
	we present the collapse-time $t_{\rm col}$ distribution of the NS for Cases~III-IV,
	where the initial angular momentum of the newborn NS is set at the mass shedding limit.
	In the left panel, the gray and blue lines are the results from the situation without and with considering an accretion disk, respectively.
	Here, the initial surface magnetic field of the newborn NS is set as $B_{\rm s,i}=10^{16}$~G.
	One can find that the $t_{\rm col}$-distributions
	are very different for the situations with and without an accretion disk.
	The $t_{\rm col}$-distribution peaks around $300$~s
	for the situations with an accretion disk,
	but peaks at around $1-10$~s for the situation without an accretion disk.
	In the right panel of Figure~\ref{fig2:time}, we show the $t_{\rm col}$-distribution
	for the situation with an accretion disk and different $B_{{\rm s},i}$, i.e., $B_{{\rm s},i}=10^{15}$~G (red lines) and $B_{{\rm s},i}=10^{16}$~G (blue lines).
	Here, the dashed, solid, and dotted lines are for the EoSs of DD2, ENG, and MPA1, respectively.
	One can find that the initial surface magnetic field strength $B_{{\rm s},i}$
	has little effect on the $t_{\rm col}$-distribution for $t_{\rm col}>10$~s,
	i.e., the peaks around $300-1500$~s.
	This is very different from the situation without considering the accretion disk,
	in which the $B_{{\rm s},i}$ makes significant effect on the $t_{\rm col}$-distribution (e.g., \citealp{2021ApJ...920..109B}).
	In addition, it seems that the $B_{{\rm s},i}$ can affect the $t_{\rm col}$-distribution for $t_{\rm col}\lesssim10$~s.
	This may reveal that the NS with short collapse-time $t_{\rm col}$
	may suffer from strong MD radiation loss before its collapse.
	According to the left panel of Figure~\ref{fig2:time},
	the peaks of the $t_{\rm col}$-distribution
	is around $\sim10^{3.3}$, $\sim10^{2.5}$, and $\sim10^{3.5}$~s
	for the EoSs of DD2, ENG, and MPA1, respectively.
	The collapse time distribution constrained by \cite{2020PhRvD.101f3021S} from the entire sample of SGRB internal plateaus shows the corresponding collapse time should be around $10^{2-3}$~s, which is more consistent with result based on the EoS of ENG. 
	
	\emph{Energy output distribution for MD radiation.\;\,}
	Except the collapse-time of a NS,
	the MD radiation energy output before its collapse is also concerned.
	In the left panel of Figure~\ref{fig3:energy},
	we plot the distribution of the total MD energy output $E_{\rm md}$ for a NS before its collapse.
	Here, the dashed, solid, and dotted lines are respectively for the EoSs of DD2, ENG, and MPA1,
	and $B_{{\rm s},i}=10^{15}$~G (red lines) or $10^{16}$~G (blue lines) is set.
	Hereafter, the situation with an accretion disk is discussed.
	One can find that the total MD energy output of a NS before its collapse depends on the initial surface magnetic field\footnote{This is different from the situation without considering an accretion disk, of which the total MD energy output is almost the same by adopting different initial surface magnetic field strength of the newborn NS (\citealp{2021ApJ...920..109B}). The significant difference between the situation with or without accretion disk is due to the burying of the magnetic field from the accreted mass, in which accretion material buries the magnetic field of NS and thus declines the MD energy output.}.
	The stronger of the initial magnetic field is,
	the higher energy output of the MD radiation would be.
	In the right panel of Figure~\ref{fig3:energy},
	we also plot the distribution of $L_{\rm md}(t_{\rm col})$.
	It reveals that the MD radiation luminosity at the collapse-time or during the internal plateau is also
	positive related to $B_{\rm s,i}$.
	According to Figure~\ref{fig3:energy},
	one have the following results.
	In the situation with $B_{{\rm s},i}\sim 10^{15}$~G,
	$E_{\rm md}\sim 10^{48}-10^{50}$~erg and $L_{\rm md}(t_{\rm col})\sim 10^{44}-10^{47}$~erg is for the EoS of DD2,
	$E_{\rm md}\sim 10^{48}-10^{50}$~erg and $L_{\rm md}(t_{\rm col})\sim 10^{44.5}-10^{47.5}$~erg is for the EoS of ENG,
	and $E_{\rm md}\sim 10^{49}-10^{51}$~erg and $L_{\rm md}(t_{\rm col})\sim 10^{43.5}-10^{47.5}$~erg is for the EoS of MPA1.
	In the situation with $B_{{\rm s},i}\sim 10^{16}$~G,
	$E_{\rm md}\sim 10^{50.5}-10^{51.7}$~erg and $L_{\rm md}(t_{\rm col})\sim 10^{45}-10^{48.5}$~erg is for the EoS of DD2,
	$E_{\rm md}\sim 10^{50}-10^{52}$~erg and $L_{\rm md}(t_{\rm col})\sim 10^{46}-10^{48.8}$~erg is for the EoS of ENG,
	and $E_{\rm md}\sim 10^{51}-10^{52.5}$~erg and $L_{\rm md}(t_{\rm col})\sim 10^{45}-10^{48.5}$~erg is for the EoS of MPA1.

	We recompile the steep decay time and the corresponding luminosity of the internal plateau
	for the sGRBs,
	which is plotted in Figure~\ref{fig4:scatter} with ``$\circ$'' symbols.
	Here, the samples with steep decay time (i.e., collapse-time $t_{\rm col}$) greater than 10~s are taken into account,
	and the solid (hollow) red ``$\circ$'' represents the data from the sGRBs with known (unknown) redshift.
	Based on our model in Section~\ref{sec:Method},
 we also plot the distribution of $L_{\rm{xrt}}-t_{\rm{col}}$ with blue contour for the NS remnants suffering from collapse (only NS remnants which survive longer than 10 s are included) in our sample of $10^5$ NS-NS mergers. Here, the initial magnetic field $B_{{\rm s},i}$ of a NS remnant is randomly selected from a normal distribution with mean $\overline{B_{{\rm s},i}}=10^{15.5}$~G ($10^{14.5}$~G) and and standard deviation $\sigma_{B_{{\rm s},i}}=0.5$,
and the left, middle, and right panels are for the situation with the EoSs of DD2, ENG, and MPA1, respectively. The upper panels correspond to the results with $B_{{\rm s},i}\sim N(\mu_{\rm B_{{\rm s},i}}=10^{15.5}$~G, $\sigma_{B_{{\rm s},i}}=0.5)$, the lower panels correspond to the results with $B_{{\rm s},i}\sim N(\mu_{\rm B_{{\rm s},i}}=10^{14.5}$~G, $\sigma_{B_{{\rm s},i}}=0.5$). As shown in Figure \ref{fig4:scatter}, the initial surface magnetic field $B_{{\rm s},i}$ only affects the distribution of  $L_{\rm{xrt}}$, while almost unaffects the distribution of $t_{\rm{col}}$. The  distribution of $L_{\rm{xrt}}$ for our samples with $B_{{\rm s},i}\sim N(\mu_{\rm B_{{\rm s},i}}=10^{15.5}$~G, $\sigma_{B_{{\rm s},i}}=0.5)$ aligns better with the observations.
	Besides, Figure~\ref{fig4:scatter} reveals that the situation with the EoS of ENG is more consistent with the observations.
	The maximum mass indicated by ENG EoS of NS satisfies $M_{\rm{max}}(P_{\rm{s}})=2.24M_\odot(1+3.95\times10^{-10} P_{\rm{s}}^{-2.26})$ and the maximum non-rotating mass $M_{\rm TOV}=2.24M_\odot$ (\citealp{2001ApJ...550..426L}).
	Based on the EoS of ENG and the mass distribution of NS-NS systems, we find the fractions of Case I, II, III, and IV are $36.5\%$, $0.7\%$, $60.7\%$, and $2.1\%$, respectively. This means over $99\%$ of NS remnants in our sample from the merger are likely to undergo a collapse process within $10^7$~s, suggesting that the majority of currently observed NSs are less likely to originate from binary NS merger remnants.

 We have assumed that the remnant begins maximally rotating (at the mass-shedding limit) in previous calculation. However, \cite{2020GReGr..52..108B} suggests that the initial spin-period of a neutron star at the beginning of the supramassive stage may not be the mass-shedding limit due to gravitational-wave emission at the earlier hypermassive stage. Therefore, we explore the distribution of collapse time (left panel) and energy output of MD radiation (right panel) with $B_{{\rm s},i}=10^{15.5}$~G in Figure \ref{fig5:different1}. 
The red, blue and green color represent the situation with $P_{{\rm s},i}=P_{{\rm max}}$, $P_{{\rm s},i}=1.2P_{{\rm max}}$ and $P_{{\rm s},i}=1.5P_{{\rm max}}$, respectively. The dashed line, solid line, and dotted line correspond to the EoSs of DD2, ENG, and MPA1, respectively.
  On the one hand, the distribution of collapse time shifts to the right, and the proportion of samples with collapse time greater than 10~s decreases significantly as $P_{{\rm s},i}$ increases for all of the EoSs. On the other hand, the distribution of MD energy shifts to the left as $P_{{\rm s},i}$ increases, which means the larger initial period causes a lower energy of MD radiation. Meanwhile, in our samples, the proportion of Case~I increases significantly, but the proportion of Case~III-IV decreases significantly with the $P_{{\rm s},i}$ increase. An excessively large initial period would result in a collapse time for NS remnants that is too long for those with a survival time exceeding 10 seconds, and the proportion of Case III—IV is too small to correspond with the observed short GRBs featuring an internal plateau. Therefore, most of the newborn NS remnants should be close to the mass-shedding limit.
	
	\section{Discussion and Conclusion} \label{sec:DISCUSS}
	The NS-NS merger can form different productions of the newborn compact remnant as follows:
	(i) the prompt formation of a black hole; (ii) the formation of a hypermassive NS collapsing to a BH in $\lesssim$1~s;
	(iii) the formation of a SMNS collapsing to a BH, or (iv) the formation of a stable NS.
	A relativistic jet may also be launched from around the compact remnant
	and is observed as a sGRB if it moves along the light of sight.
	Since the formed SMNS could collapse into a BH,
	an internal plateau may appear in the light-curve of sGRB's afterglow.
	Then, the sGRB with an internal plateau is generally used to represent the formation of a SMNS.
	The formation fraction and the collapse-time of SMNSs are generally used to constrain the EoS of NS
	(e.g., \citealp{2014PhRvD..89d7302L}; \citealp{2016PhRvD..93d4065G}).
	For this topic, however,
	the sGRBs with an internal plateau is only related to the production of (iii) in previous works.
	It is worth pointing out that a significant ejecta is expelled during the NS-NS merger
	and subsequently falls back onto the compact remnant by accretion.
	Some stable NS may transform into a SMNS or BH during this process.
	That is to say, the sGRBs with an internal plateau is related to the production of both (iii) and (iv),
	rather than only the production of (iii).
	In this paper, we revisit the topic of the constraint on the NSs' EoS,
	based on the Cases III-IV by considering an accretion disk around the compact remnant for the NS-NS mergers.

	Since the internal plateau may be missed if the external shock emission is dominant in the afterglow,
	the proportion of both (iii) and (iv) in the four productions should be higher than
	the proportion of sGRBs with an internal plateau.
	It is found that only the situation with three kinds of EoS,
	i.e., DD2, ENG, and MPA1,
	is consistent with the observations.
	With an accretion disk around the compact remnant,
	the collapse-time $t_{\rm col}$ of the NSs formed in the NS-NS mergers
	peaks at around $10^{3.3}$, $10^{2.5}$, and $10^{3.5}$~s
	for the EoSs of DD2, ENG, and MPA1, respectively.
	In addition, the above peaks is almost independent on the initial magnetic field of the newborn NS.
	Observationally, the steep decay time of the internal plateau
	is around $10^{2-3}$~s,
	which is in favour of ENG for the EoS of NSs.
	We also study the diagram of the steep-decay time (corresponding to the collapse-time of a SMNS)
	and the corresponding luminosity
	of the internal plateau for the sGRBs.
	It is shown that only the situation with the EoS of ENG is well consistent with the observations, which suggests the the maximum non-rotating mass $M_{\rm TOV}=2.24M_\odot$ (\citealp{2001ApJ...550..426L}).
It should be noted that the results above are based on samples which is only associated with NS-NS binaries detected in the Milky Way. However, GW~190425 whose chirp mass and total mass are significantly different from NS-NS binaries detected in the Milky Way is excluded. If GW~190425 is a NS-NS merger event, it is suggested that there are two kinds of event for NS-NS merger i.e. NS-NS binaries detected in the Milky Way and NS-NS binaries similar to GW~190425. Therefore, we re-sample from the double
peaked Gaussian probability distribution fitted by the full population of galactic neutron stars (\citealp{2018MNRAS.478.1377A}) 
\begin{equation}
 P(M)=(1-\epsilon)\mathcal{N}(\mu_{\rm 1},\sigma_{\rm 1})+\epsilon\mathcal{N}(\mu_{\rm 2},\sigma_{\rm 2}),
\end{equation}
with $\mu_{\rm 1}=1.32M_{\odot}, \sigma_{\rm 1}=0.11, \mu_{\rm 1}=1.80M_{\odot}, \sigma_{\rm 1}=0.21$ and mixing fraction $\epsilon=0.35$. The majority of observed NS binary systems (include GW~190425) do not exhibit extreme mass ratios, which means the NSs in $\mathcal{N}(\mu_{\rm 1},\sigma_{\rm 1})$ are more likely to merger with the NSs in $\mathcal{N}(\mu_{\rm 1},\sigma_{\rm 1})$ and the NSs in $\mathcal{N}(\mu_{\rm 2},\sigma_{\rm 2})$ are more likely to merger with the NSs in $\mathcal{N}(\mu_{\rm 2},\sigma_{\rm 2})$. We re-plot Figure \ref{fig1:EOS} and find that it does not affect our selection for EoSs (DD2, ENG, and MPA1) initially, and the majority of samples forming case III-IV whose remnants surviving longer than 10~s only correspond to NS-NS systems from $\mathcal{N}(\mu_{\rm 1},\sigma_{\rm 1})$, the vast majority of the NS-NS systems from $\mathcal{N}(\mu_{\rm 2},\sigma_{\rm 2})$ form BHs promptly or collapse within 10~s. This means the distribution for collapse time and MD energy output for samples whose remnants surviving longer than 10~s is affected by the double peaked Gaussian probability distribution very slightly, and most of SGRBs with internal plateaus should origin from the merger of NS-NS from $\mathcal{N}(\mu_{\rm 1},\sigma_{\rm 1})$. A larger mixing fraction $\epsilon$ causes higher fraction for Case~I. Therefore, the EoS of ENG is still more favored.
 Previous studies have also constrained the maximum non-rotating mass $M_{\rm TOV}$ of the NSs. \cite{2024PhRvD.109d3052F} has suggested $M_{\rm TOV}=2.25^{+0.08}_{-0.07}M_\odot$ by analyzing the mass distribution of all observed NS and reconstructing the EoS of the very dense matter. \cite{2020PhRvD.101f3021S} has presented $M_{\rm TOV}=2.26^{+0.31}_{-0.17}M_\odot$ assuming a variable braking index and GW~190425 is not a binary NS merger, $M_{\rm TOV}=2.31^{+0.36}_{-0.21}M_\odot$ when GW~190425 is considered to be a NS-NS merger. Coincidentally, our result is close to these previous studies.
	Since various conditions and parameters may influence our results, it is necessary to explore our results under different conditions, which includes the mass of the unbound wind ejecta ($M_{\rm{eject,w}}$) and the angle between the magnetic-field and spin-axes ($\chi$). Based on the EoS of ENG, we explore our results under different conditions. In this paper, we adopt a constant fraction $M_{\rm{eject,w}} \approx 0.2M_{\rm{disk,total}}$ of the disk for unbound wind ejecta. However, numerous numerical simulations suggest that longer neutron star survival times likely unbind a bigger fraction of the disk as a result of additional neutrino cooling from the neutron star (\citealp{2021ApJ...906...98N,2019ApJ...886L..30N,2018ApJ...860...64F,2017ApJ...846..114F}). Therefore, we plot the probability density distribution of the collapse time (left panel) and energy output of the MD radiation (right panel) for the samples of the case~III-IV under $M_{\rm{eject,w}} \approx 0.2M_{\rm{disk,total}}$ (solid line), $M_{\rm{eject,w}} \approx 0.4M_{\rm{disk,total}}$ (dashed line), $M_{\rm{eject,w}} \approx 0.6M_{\rm{disk,total}}$ (dotted line) with $B_{{\rm s},i}=10^{15}$~G (red) and $B_{{\rm s},i}=10^{16}$~G (blue) in the upper panel of the Figure \ref{fig5:different}.
			For $B_{{\rm s},i}=10^{15}$~G, an increase in $M_{\rm{eject,w}}$ leaves the distribution of collapse time largely unaffected, while the energy output of the MD radiation increases. Conversely, for $B_{{\rm s},i}=10^{16}$~G, as $M_{\rm{eject,w}}$ increases, the distribution of collapse time shifts to the right, and the proportion of samples with a collapse time greater than 10 s decreases, accompanied by an increase in the energy output of the MD radiation. Besides, as long as the wind mass loss does not exceed $0.4M_{\rm disk}$ too much, the the wind mass loss has little influence on our results.

Beside, the angle between the magnetic-field and spin-axes, denoted as $\chi$, is another factor that may influence our results, as it affects the torque exerted by gravitational waves on the neutron star remnant. The torque exerted by gravitational waves is sensitive to the angle between the magnetic-field and spin-axes (\citealp{2002PhRvD..66h4025C}) which is subject to significant evolution (\citealp{2020MNRAS.494.4838L}). Therefore, we change the angle between the magnetic-field and spin-axes to explore the corresponding influence. We plot the distribution of collapse time (left panel) and energy output of MD radiation (right panel) with $\chi=\rm\pi/2$ (solid line), $\chi=\rm\pi/4$ (dashed line), $\chi=0$ (dotted line) in the lower panel of Figure \ref{fig5:different}. As shown in the lower panel of Figure \ref{fig5:different}, the distribution of collapse time for samples with a collapse time greater than 10~s slightly shifts to the right as $\chi$ decreases (the GW radiation decreases but MD radiation increase). Overall, the distribution of collapse time and energy output of MD radiation is minimally affected by the angle between the magnetic-field and spin-axes. In addition, it's probable that various GW emission processes operate in the early life of the neutron star, such as the r-mode instability (\citealp{2001IJMPD..10..381A}) and the secular bar-mode instability (\citealp{2003CQGra..20R.105A}). The saturation amplitudes of the secular bar mode and inertial r mode are quite uncertain. Considering that the gravitational wave (GW) radiation from the magnetic deformation of the remnants may be relatively strong compared to the r-mode instability (\citealp{2018ASSL..457..673G,2023RAA....23b5001C}), it is believed that the GW radiation from the r-mode instability will have a minor effect on our results. For the secular bar-mode instability, all of our samples during their evolution progress show a lower ratio of the rotational kinetic energy $T$ to gravitational binding energy $W$ than 0.14, i.e. $T/|W|<0.14$, which means the secular bar-mode instability is ineffective for our samples.

	The further discussion based on the EoS of ENG for the NS are presented as follows.
	In Figure~\ref{fig7:colorbar2},
	we plot the $M_{\rm{s}}-M_{\rm{disk}}$ ($M_{\rm{chirp}}-M_{\rm{disk}}$) distributions for the Case III-IV in our sample at the mass-shedding limit,
	where the value of $t_{\rm col}$ is showed with color bar.
	By comparing the left panel ($B_{{\rm s},i}=10^{15}$~G) and the right panel ($B_{{\rm s},i}=10^{16}$~G) of this figure,
	one can find that the initial surface magnetic field strength $B_{{\rm s},i}$ of the newborn NS
	only makes its effect for the NS remnant with high mass.
	The large mass of the disk or the NS remnant for Case~III-IV would have a short collapse-time for the NS.
	Besides, the mass range of NS from Cases III-IV is $2.1-2.51 M_{\odot}$
	and the chirp mass of the corresponding NS-NS binary is between $1.04-1.2M_{\odot}$.
	Importantly, if the collapse-time of the NS remnant is in the range of $10-10^3$~s,
	the mass of the newborn NS should be at around $2.21-2.49 M_{\odot}$
	and the chirp mass of the corresponding NS-NS binary is $1.085-1.182 M_{\odot}$ for $B_{{\rm s},i}\sim 10^{15}-10^{16}$~G.
	In the following, we discuss three gravitational-wave events based on the results in Figure~\ref{fig7:colorbar2}.
	(1) GW~170817.\,
	It is shown that the total mass of the NS binary system for GW~170817 is 2.74~$M_{\odot}$ (\citealp{2017PhRvL.119p1101A}).
	The ejecta from this NS-NS merger is estimated at around $0.04$~$M_{\odot}$ (e.g., \citealp{2017Natur.551...75S,2019ApJ...885...60R}).
	Then, it can be believed that a black hole should be formed within 1~s after GW~170817,
	unless the mass loss through gravitational-wave for this merger exceeds 0.3~$M_{\odot}$ significantly. Some previous studies also favor a BH for GW~170817 (\citealp{2017ApJ...850L..24G,2017ApJ...850L..19M,2018ApJ...854...60M,2021ApJ...920..109B}).
	(2) GW~190425.\,
	The total mass of the corresponding binary for this event is estimated at $\sim 3.4\,M_{\odot}$ \citep{2020ApJ...892L...3A}.
	Then, it can believe that the compact remnant of this merger should be a black hole, unless the total mass of accretion disk, dynamic ejecta and the loss by gravitational-wave exceeds 0.9~$M_{\odot}$ significantly. Some previous studies also favor a BH for GW~190425 (\citealp{2020ApJ...892L...3A,2021GReGr..53...59S}).
	(3) GW~190814.\,
	In GW~190814, the estimated mass of lighter compact stars is around $M\sim(2.5-2.67)\,M_{\odot}$.
	Consensus has not been achieved on whether the secondary object is a neutron star or a black hole (\citealp{2021PhRvL.126p2702B,2021MNRAS.505.1600B}). The maximum non-rotating mass of a NS is 2.24~$M_{\odot}$ and the maximum mass of a NS is 2.51~$M_{\odot}$ in our samples for the ENG EoS. Then, the secondary object should be a black hole, which is similar as some other previous studies (\citealp{2020PhRvC.102f5805F,2020PhRvL.125z1104T,2021MNRAS.505.1600B,2021ApJ...908L..28N}).
	
Finally, we also plot the $M_{\rm{s}}-M_{\rm{disk}}$ distributions for Case III-IV in our sample, with $P_{{\rm s},i}=1.5P_{\rm max}$, in Figure \ref{fig8:colorbar3}. Under this condition, the collapse time distribution is affected only by $M_{\rm{s}}$ and $M_{\rm{disk}}$, and not by the initial surface magnetic field ($B_{{\rm s},i}$) of the remnants. For Case III-IV, a larger mass of the disk or the NS remnant corresponds to a shorter collapse time for the NS. Besides, the mass range of NS from Cases III-IV decrease to $2.1-2.32 M_{\odot}$.
\cite{2022ApJ...939...51M} presents three evolutionary scenarios for NS remnants with masses less than $M_{\rm TOV}$: In Case~A (for NS initially close to the mass-shedding limit and with a relatively small mass), the NS will spin up to the mass-shedding limit, continue spinning at the mass-shedding limit until accretion ceases, and then spin down before collapsing. In Case~B (for NS with large initial angular momentum and a relatively large mass), the NS will spin up until accretion stops, and then spin down before collapsing. In Case~C (for NS with small initial angular momentum and a large mass), the NS will collapse during the accretion process.
According to Figure~\ref{fig7:colorbar2}, the NS in Case~A, as presented by \cite{2022ApJ...939...51M}, takes longer than $10^3$~s to collapse. As shown in Figure~\ref{fig8:colorbar3}, the collapse time of the NS in Cases~B and C is very uncertain; some may collapse in less than 1~s, while others may take longer.

\emph{Acknowledgments}
{ We thank the anonymous referee for helpful feedback on the manuscript.
This work is supported by the National Natural Science Foundation of China
(grant Nos. 12273005 and 12133003), the Guangxi Science Foundation (grant Nos. 2018GXNSFFA281010), and China Manned Spaced Project (CMS-CSST-2021-B11).}
	
	\clearpage
	
	\clearpage
	\bibliographystyle{aasjournal}
	\bibliography{sample631}

	\begin{figure}
		\centering
		\includegraphics[width=0.7\textwidth]{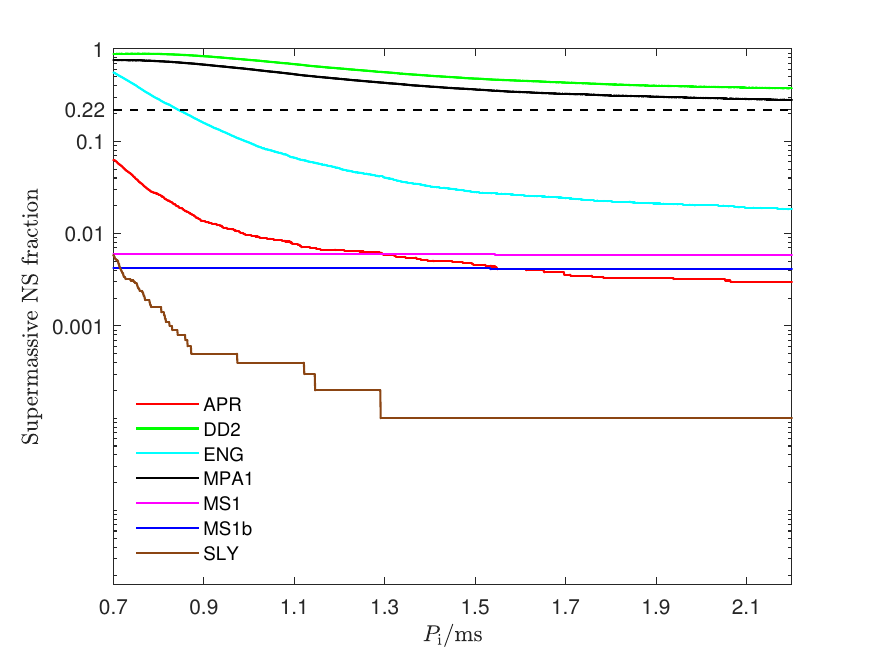}
		\caption{ Formation fraction of Cases III-IV vs. the initial period $P_{{\rm s},i}$ of the compact object,
			where the lines with different color represent the situation with different EoS.
			The horizontal dashed-line corresponds to the fraction of $22\%$,
			which is set as the lower bound for the formation fraction of Cases III-IV.}
		\label{fig1:EOS}
	\end{figure}

	\begin{figure*}[htb]
		\centering
		\begin{minipage}{0.49\linewidth}
			\includegraphics[height=0.7\textwidth, width=1\textwidth]{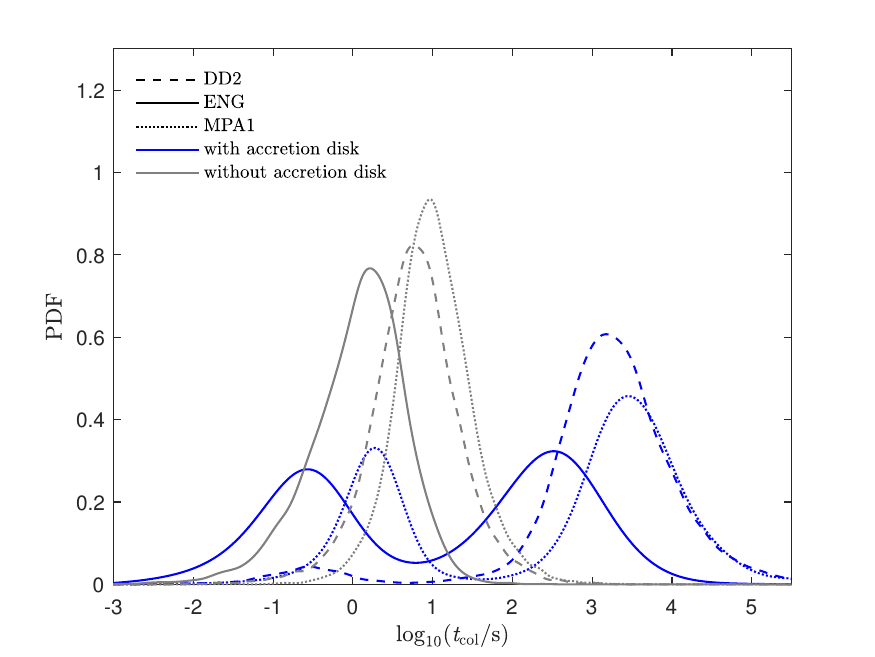}
		\end{minipage}
		\begin{minipage}{0.49\linewidth}
			\includegraphics[height=0.7\textwidth, width=1\textwidth]{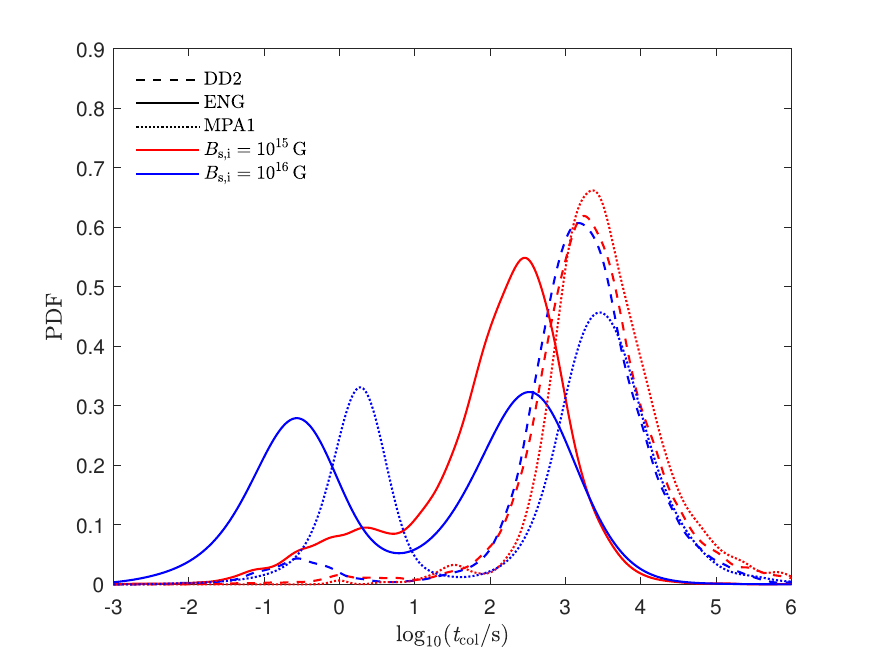}
		\end{minipage}
		
		\caption{Probability density distribution (PDF) of collapse-time ($t_{\rm{col}}$) for the Cases III-IV. The dashed line, solid line, and dotted line correspond to the EoSs of DD2, ENG, and MPA1, respectively. The red and blue color represent the situation with $B_{{\rm s},i}=10^{15}$~G and $B_{{\rm s},i}=10^{16}$~G, respectively. The gray color represents the situation without considering the accretion disk.}
		\label{fig2:time}
	\end{figure*}
	
	\begin{figure*}[htb]
		\centering
		\begin{minipage}{0.49\linewidth}
			\includegraphics[height=0.7\textwidth, width=1\textwidth]{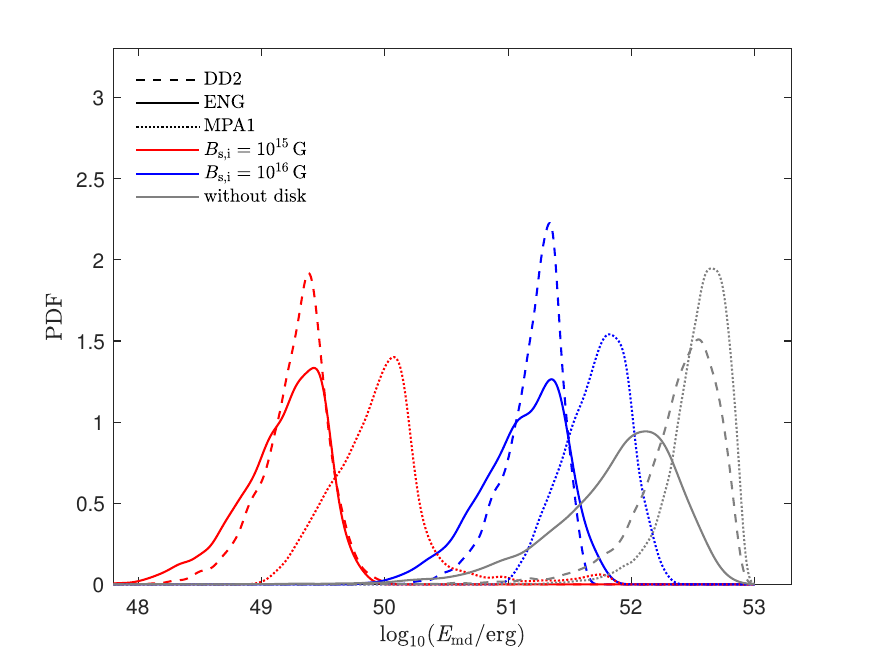}
		\end{minipage}
		\begin{minipage}{0.49\linewidth}
			\includegraphics[height=0.7\textwidth, width=1\textwidth]{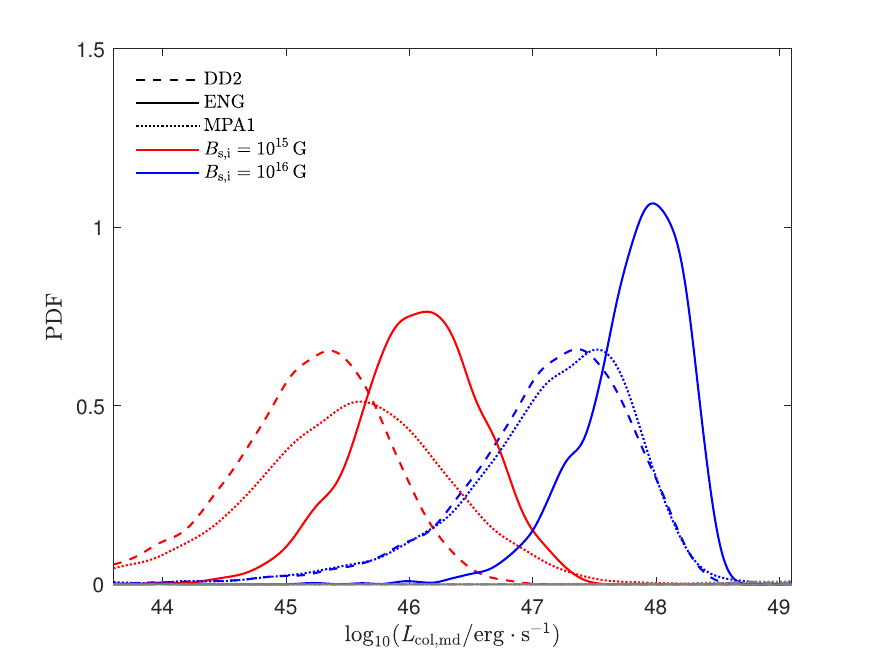}
		\end{minipage}
		
		\caption{Probability density distribution (PDF) of the total energy output of the MD radiation ($E_{\rm{md}}$) and the luminosity ($L_{\rm{col,md}}$) for a SMNS in Cases III-IV. The dashed line, solid line, and dotted line correspond to the EoSs of DD2, ENG, and MPA1, respectively. The color red and blue represent the situation with $B_{{\rm s},i}=10^{15}$~G and $B_{{\rm s},i}=10^{16}$~G, respectively.}
		\label{fig3:energy}
	\end{figure*}
	
\begin{figure*}
		\centering
		\begin{minipage}{0.32\linewidth}
			\includegraphics[height=0.9\textwidth, width=1\textwidth, keepaspectratio]{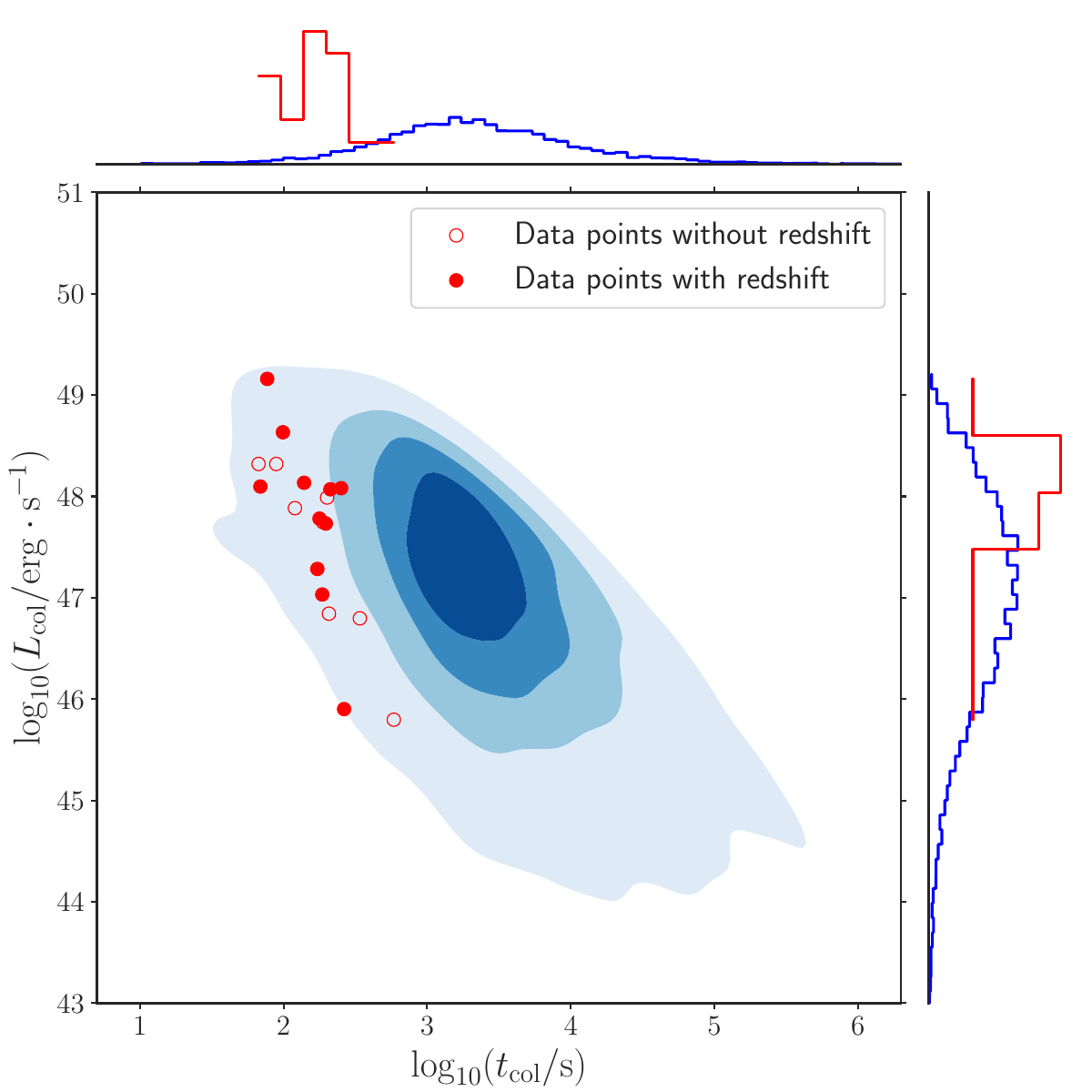}
			\label{fig:subfig1}
		\end{minipage}
		\hfill
		\begin{minipage}{0.32\linewidth}
			\includegraphics[height=0.9\textwidth, width=1\textwidth, keepaspectratio]{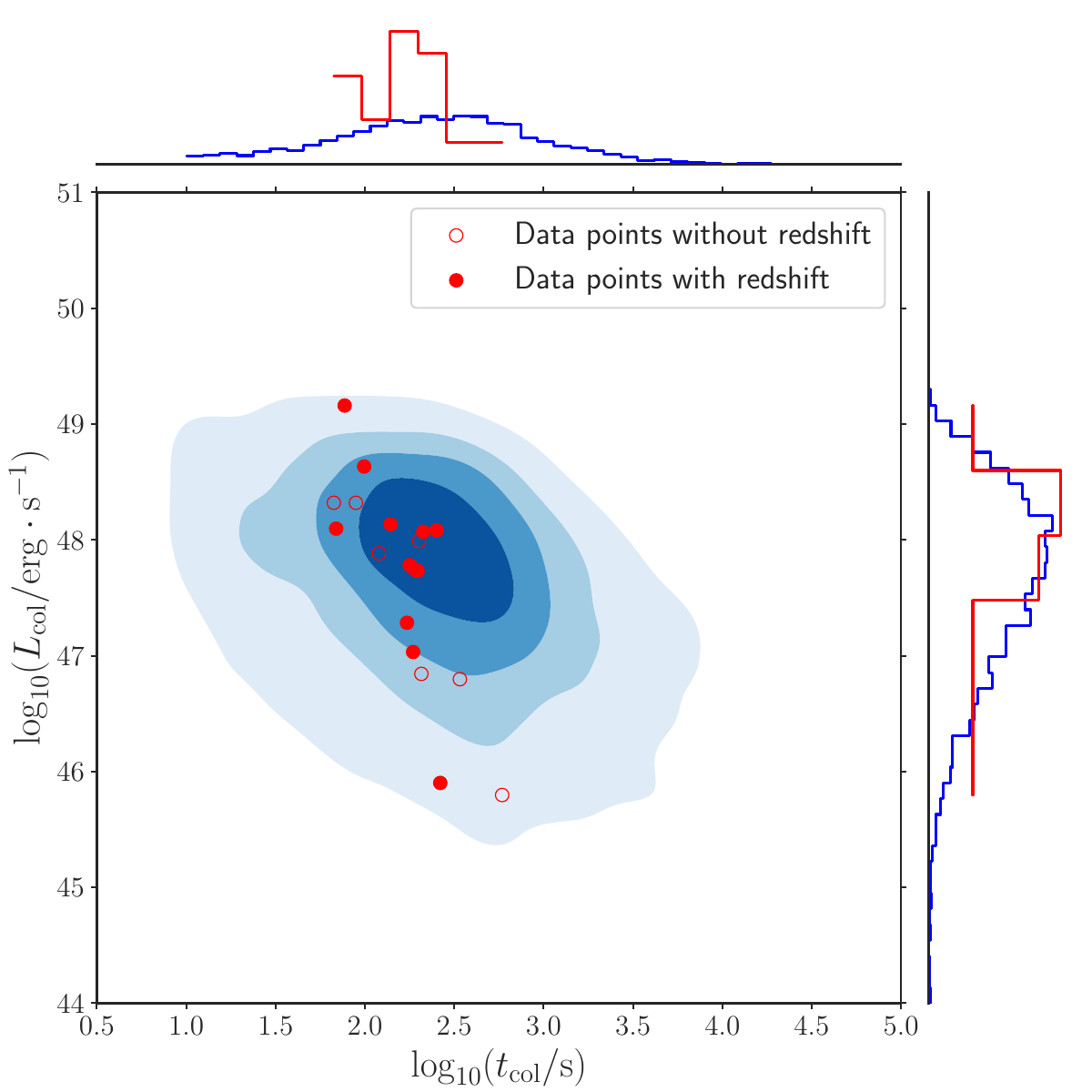}
			\label{fig:subfig2}
		\end{minipage}
		\hfill
		\begin{minipage}{0.32\linewidth}
			\includegraphics[height=0.9\textwidth, width=1\textwidth, keepaspectratio]{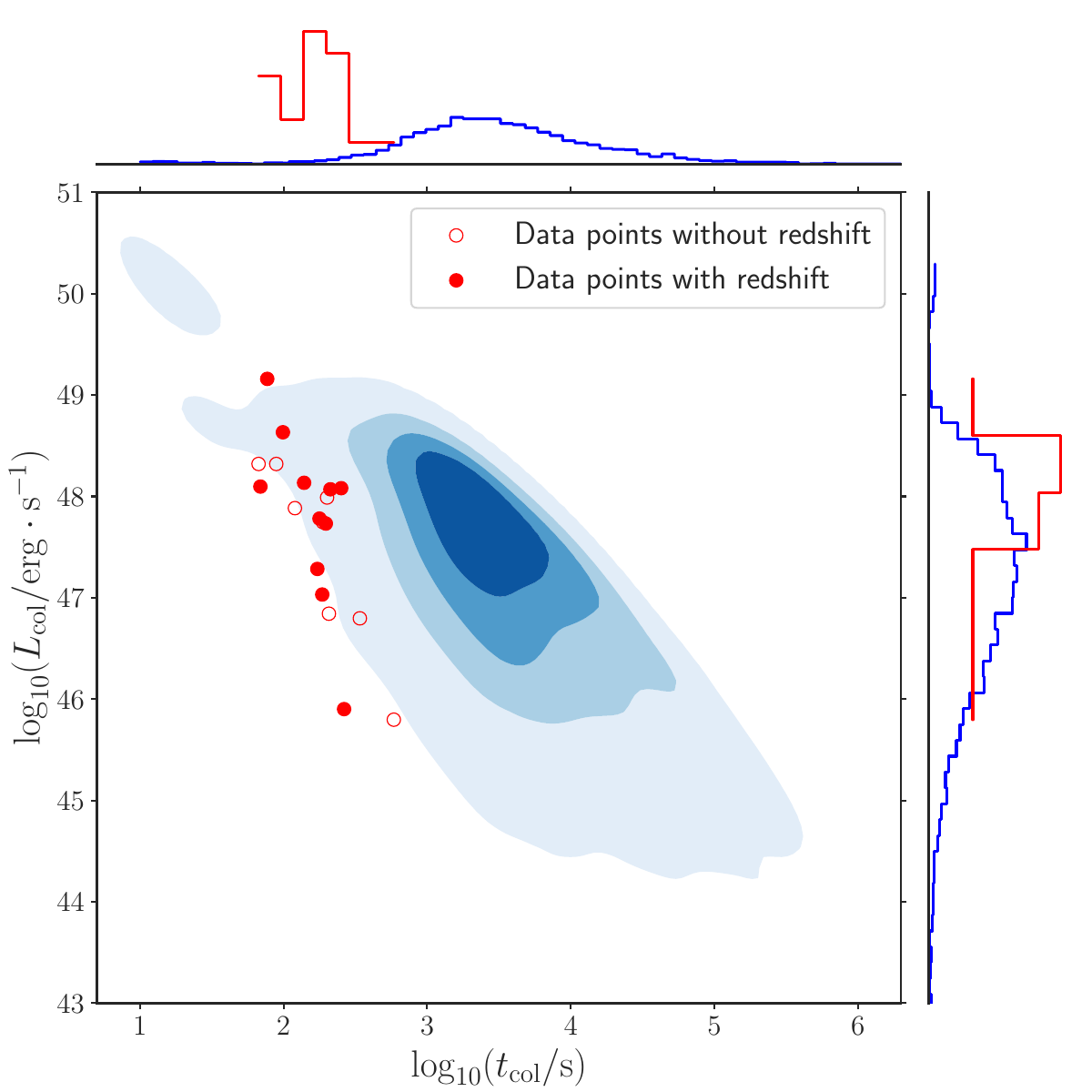}
			\label{fig:subfig3}
		\end{minipage}

  	\begin{minipage}{0.32\linewidth}
			\includegraphics[height=0.9\textwidth, width=1\textwidth, keepaspectratio]{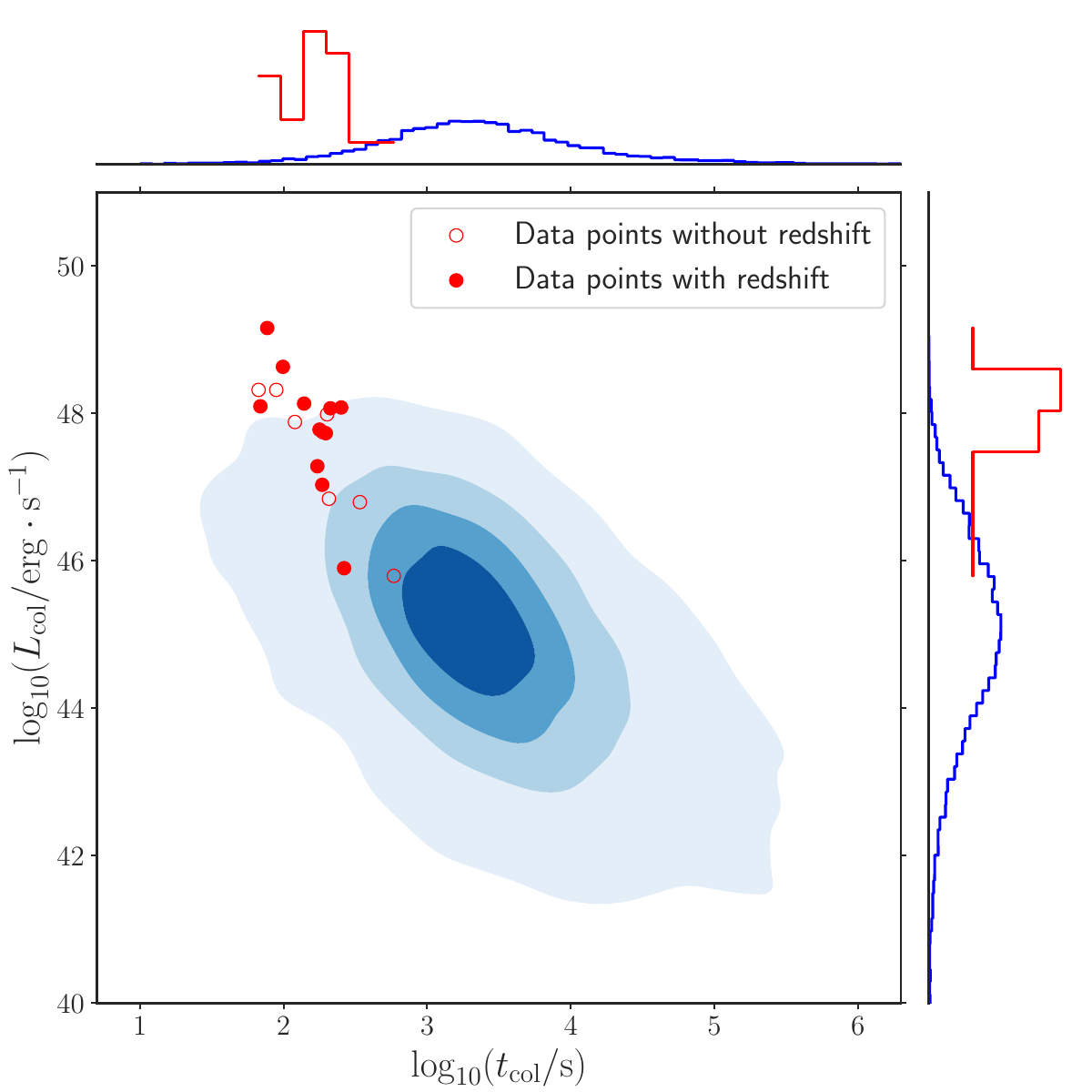}
			\label{fig:subfig4}
		\end{minipage}
		\hfill
		\begin{minipage}{0.32\linewidth}
			\includegraphics[height=0.9\textwidth, width=1\textwidth, keepaspectratio]{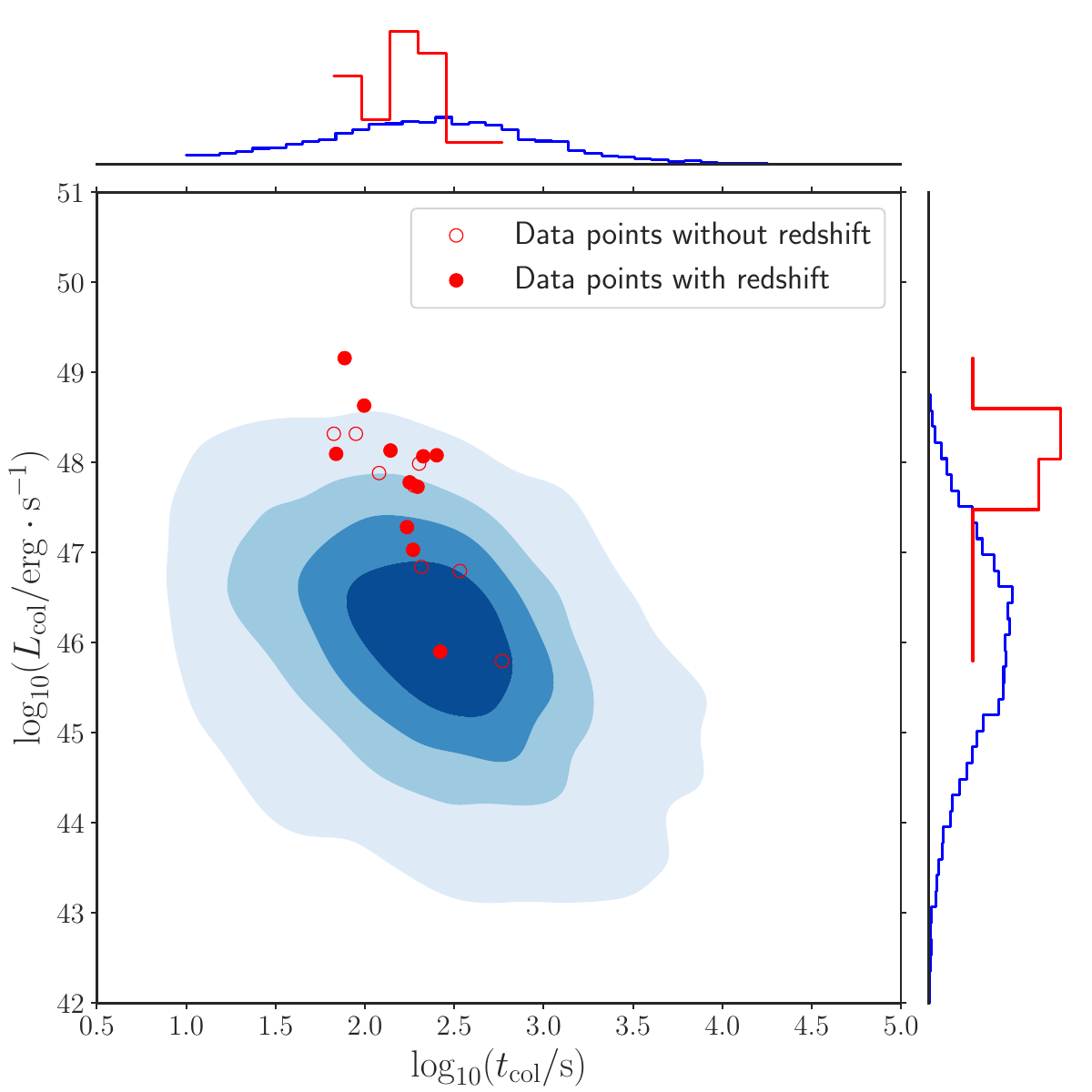}
			\label{fig:subfig5}
		\end{minipage}
		\hfill
		\begin{minipage}{0.32\linewidth}
			\includegraphics[height=0.9\textwidth, width=1\textwidth, keepaspectratio]{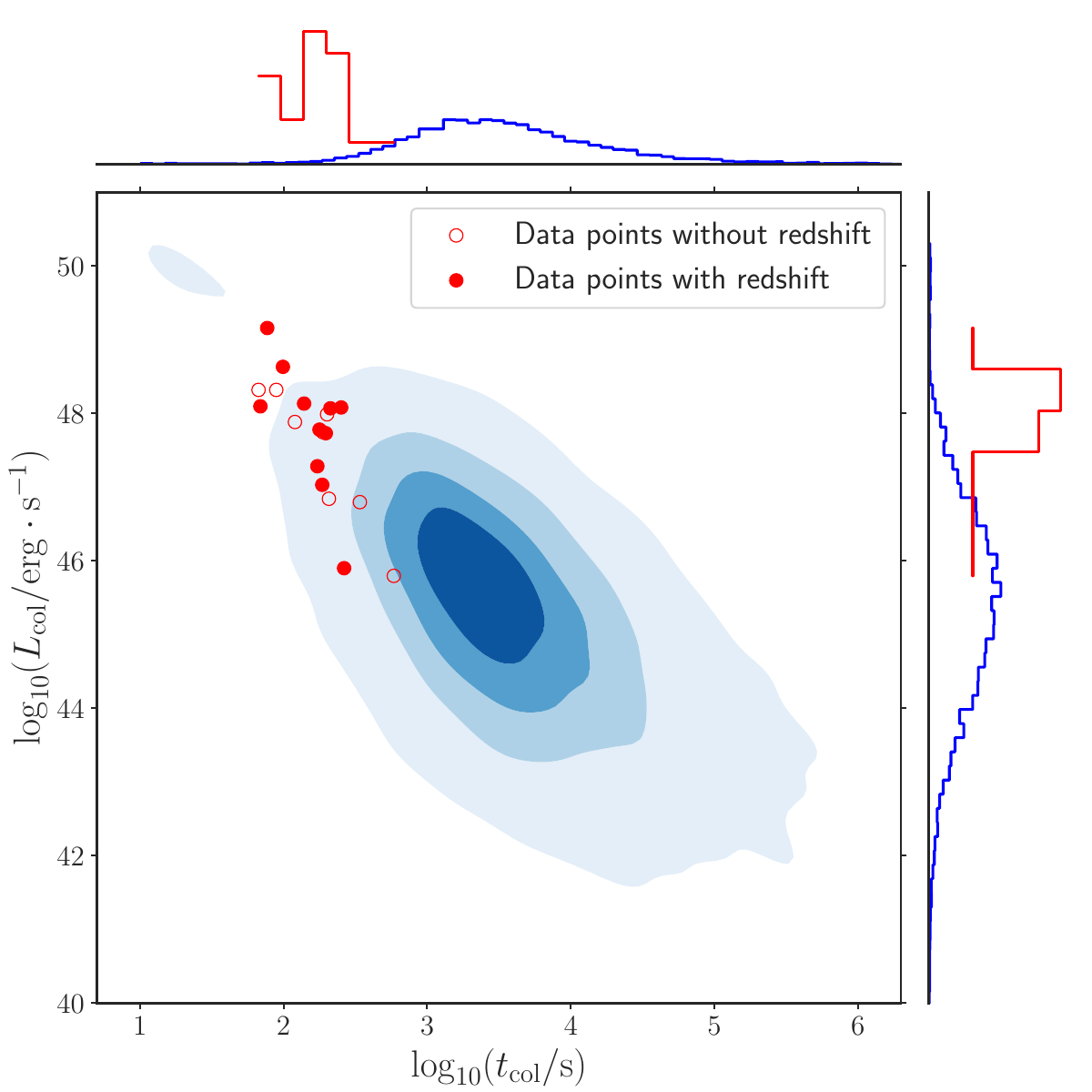}
			\label{fig:subfig6}
		\end{minipage}
		
		\caption{The probability density distribution (PDF) for the Case~III-IV (only samples with collapse time larger than 10~s are included) in the $L_{\rm{col}}-t_{\rm{col}}$ diagram,
			where the solid (hollow) red ``$\circ$'' represents the data from the sGRBs with known (unknown) redshift,
			and the left, middle, and right figures correspond to the DD2, ENG, and MPA1 EoSs, respectively. The upper panels correspond to the results with $B_{{\rm s},i}\sim N(\mu_{\rm B_{{\rm s},i}}=10^{15.5}$~G, $\sigma_{B_{{\rm s},i}}=0.5)$, the lower panels correspond to the results with $B_{{\rm s},i}\sim N(\mu_{\rm B_{{\rm s},i}}=10^{14.5}$~G, $\sigma_{B_{{\rm s},i}}=0.5$).}
		\label{fig4:scatter}
	\end{figure*}

 \begin{figure*}

		\begin{minipage}{0.49\linewidth}
			\includegraphics[height=0.9\textwidth, width=1\textwidth, keepaspectratio]{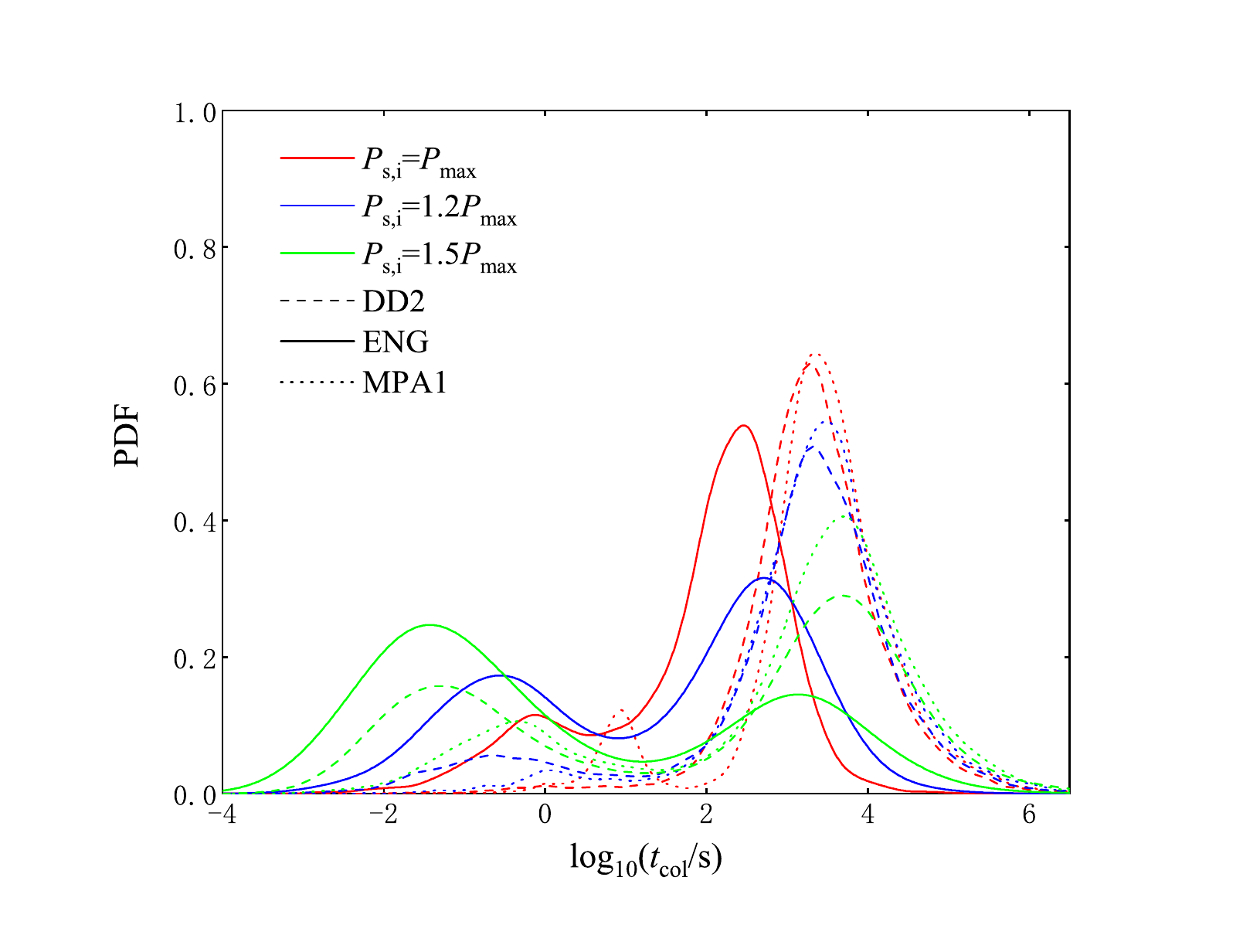}
			
		\end{minipage}
		\begin{minipage}{0.49\linewidth}
			\includegraphics[height=0.9\textwidth, width=1\textwidth, keepaspectratio]{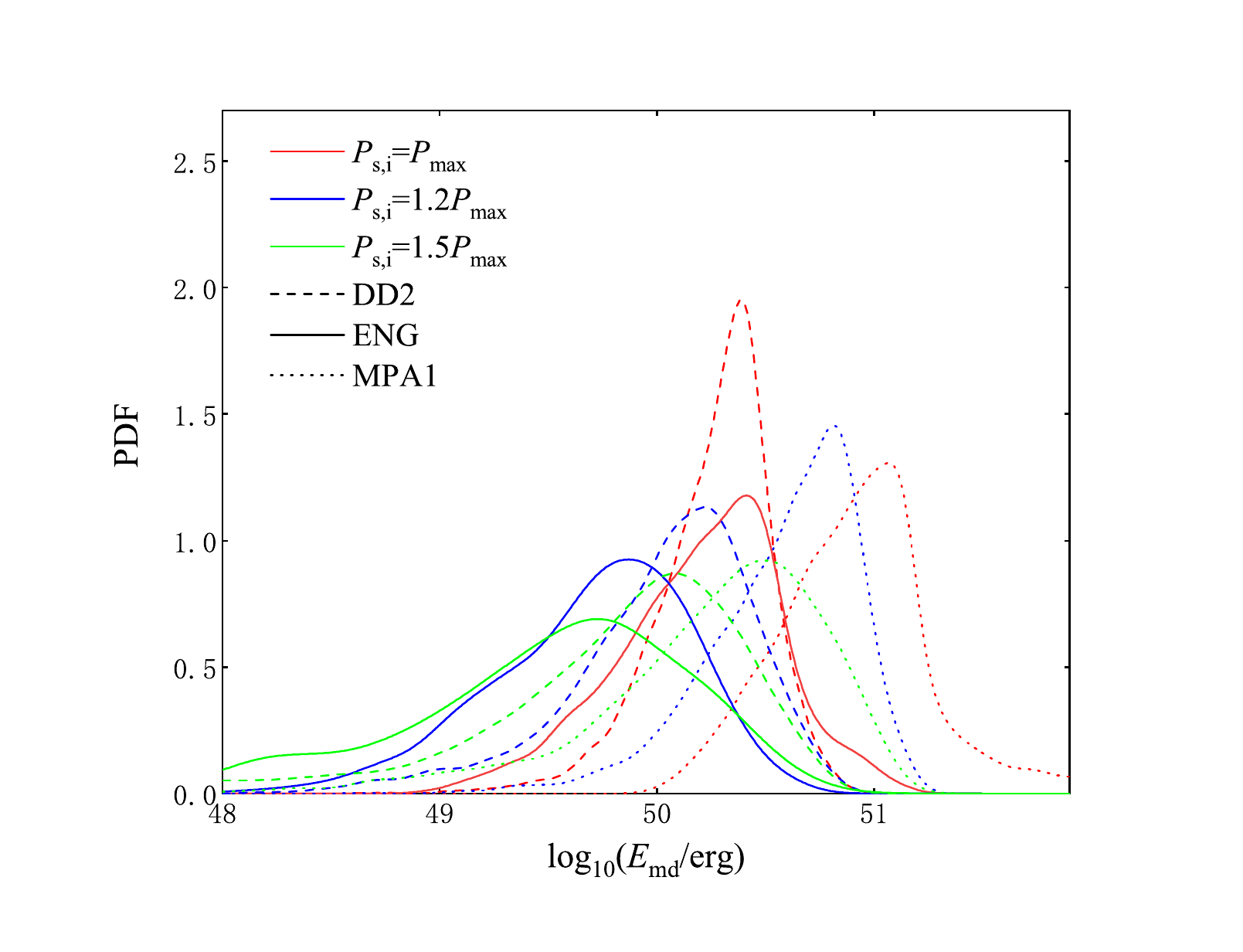}
			
		\end{minipage}	
		\caption{The probability density distribution (PDF) of the collapse-time (left panel) and energy output of the MD radiation (right panel) with $B_{{\rm s},i}=10^{15.5}$~G for Case~III-IV from our sample. The red, blue and green color represent the situation with $P_{{\rm s},i}=P_{{\rm max}}$, $P_{{\rm s},i}=1.2P_{{\rm max}}$, $P_{{\rm s},i}=1.5P_{{\rm max}}$, respectively. The dashed line, solid line, and dotted line correspond to the EoSs of DD2, ENG, and MPA1, respectively.}
		\label{fig5:different1}
	\end{figure*}
	\begin{figure*}
		\centering
		\begin{minipage}{0.49\linewidth}
			\includegraphics[height=0.9\textwidth, width=1\textwidth, keepaspectratio]{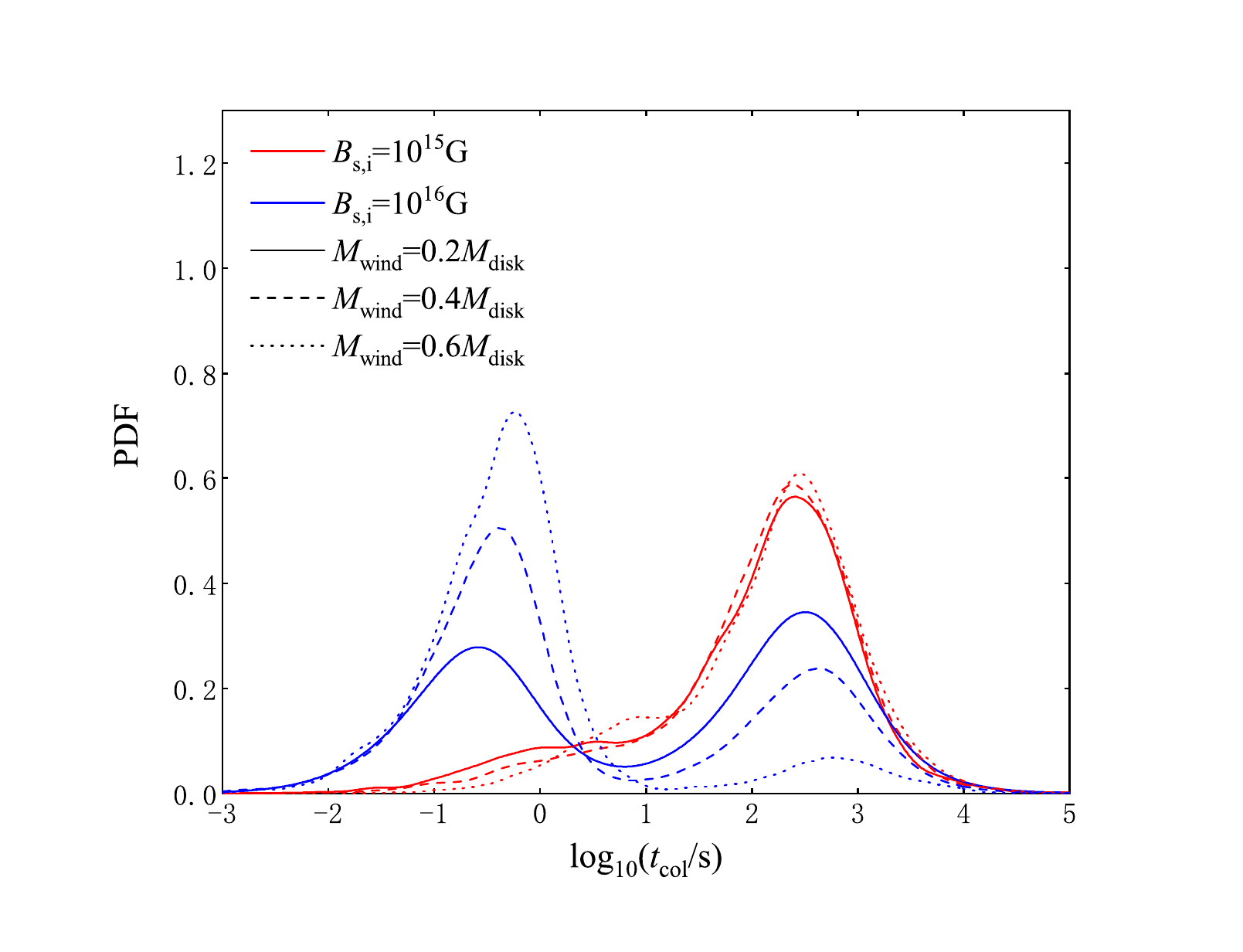}
			
		\end{minipage}
		\hfill
		\begin{minipage}{0.49\linewidth}
			\includegraphics[height=0.9\textwidth, width=1\textwidth, keepaspectratio]{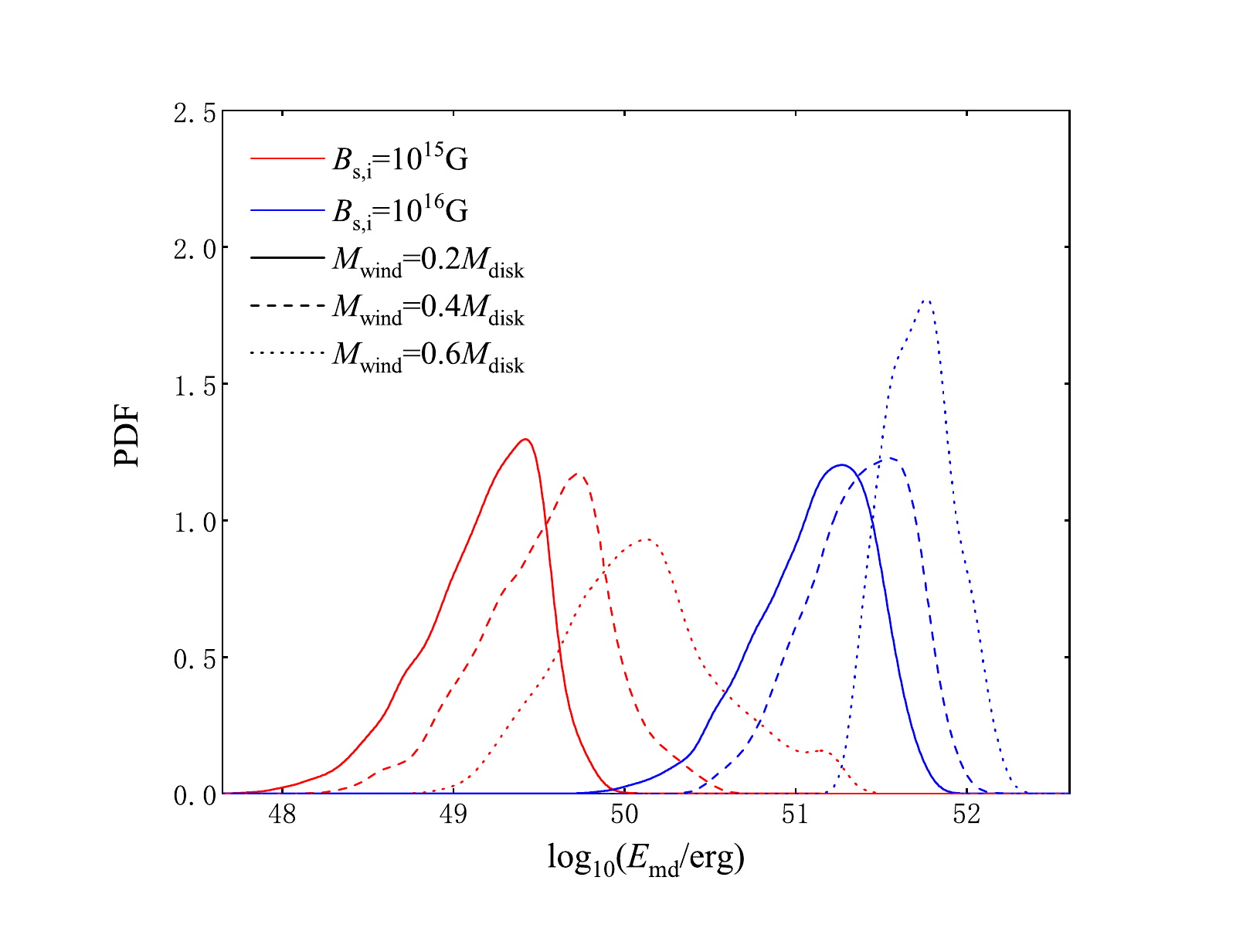}
			
		\end{minipage}
		\hfill
		
		\begin{minipage}{0.49\linewidth}
			\includegraphics[height=0.9\textwidth, width=1\textwidth, keepaspectratio]{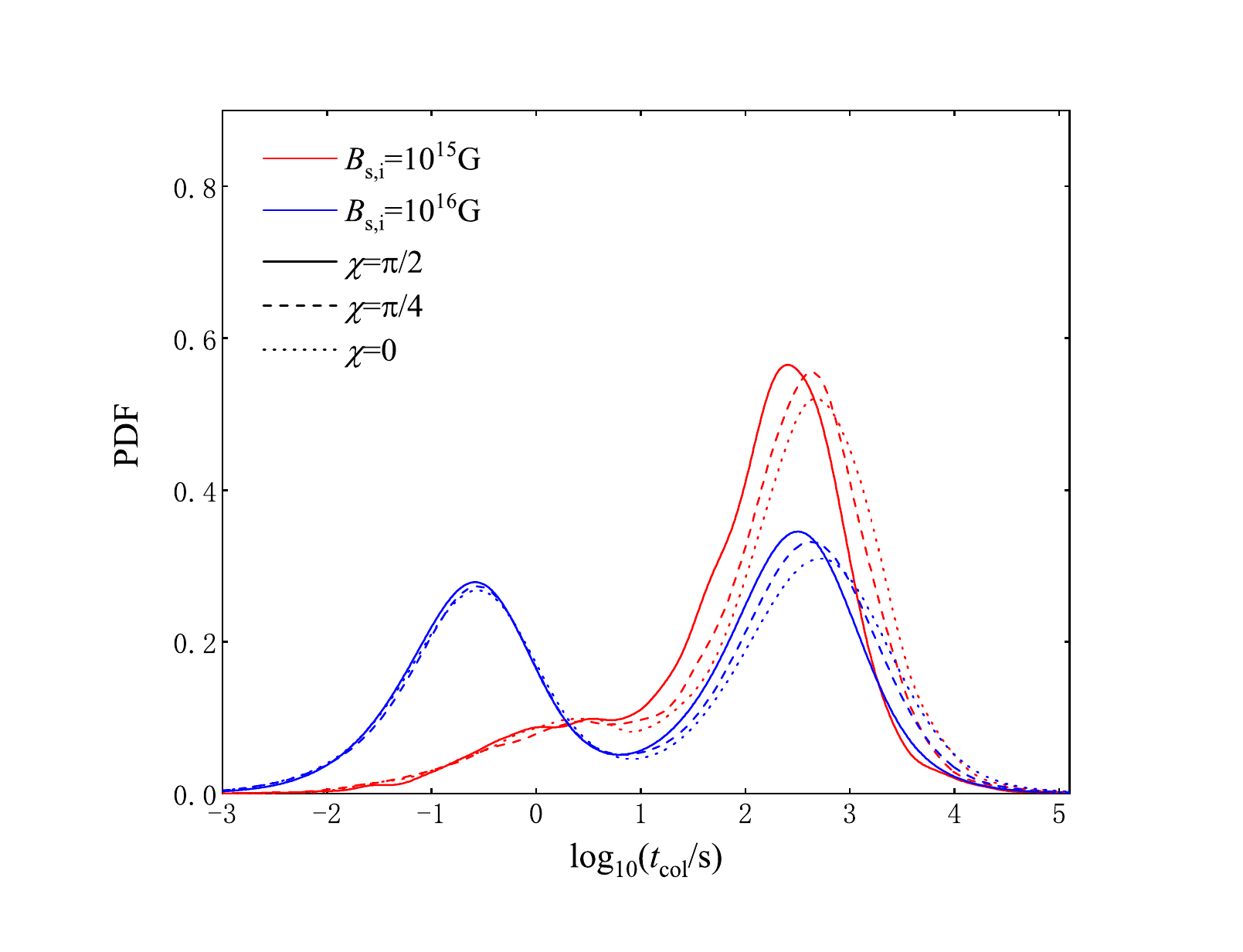}
			
		\end{minipage}
		\hfill
		\begin{minipage}{0.49\linewidth}
			\includegraphics[height=0.9\textwidth, width=1\textwidth, keepaspectratio]{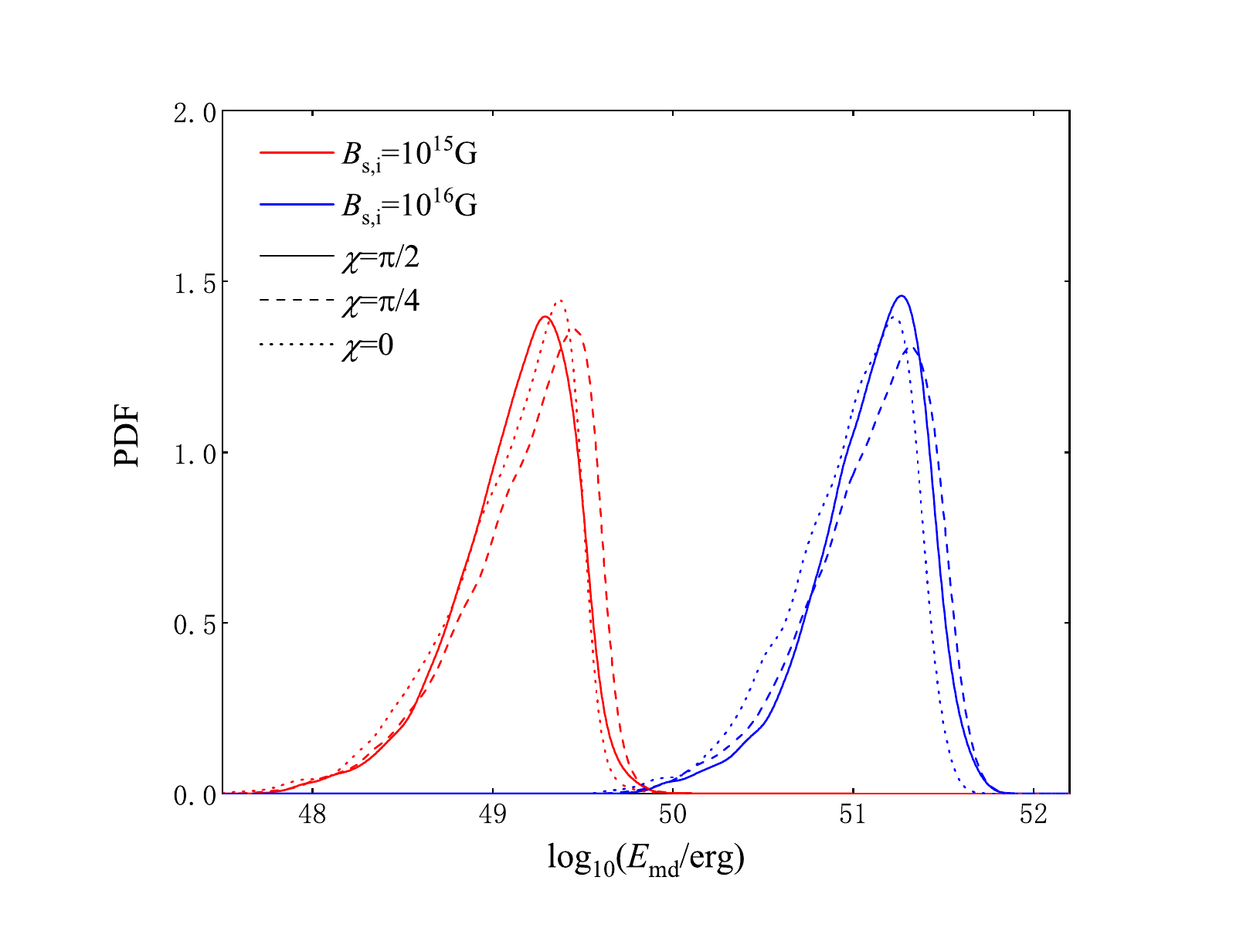}
			
		\end{minipage}	
		\caption{The probability density distribution (PDF) of the collapse-time (left panel) and energy output of the MD radiation (right panel) for Case~III-IV from our sample are presented under different conditions of $M_{\rm{wind}}$ (upper panels), and $\chi$ (lower panels). The red and blue color represent the situation with $B_{{\rm s},i}=10^{15}$~G and $B_{{\rm s},i}=10^{16}$~G, respectively. Different line styles represent different conditions.}
		\label{fig5:different}
	\end{figure*}

 \begin{figure*}[htb]
		\centering
		\begin{minipage}{0.49\linewidth}
			\includegraphics[height=0.7\textwidth, width=1\textwidth]{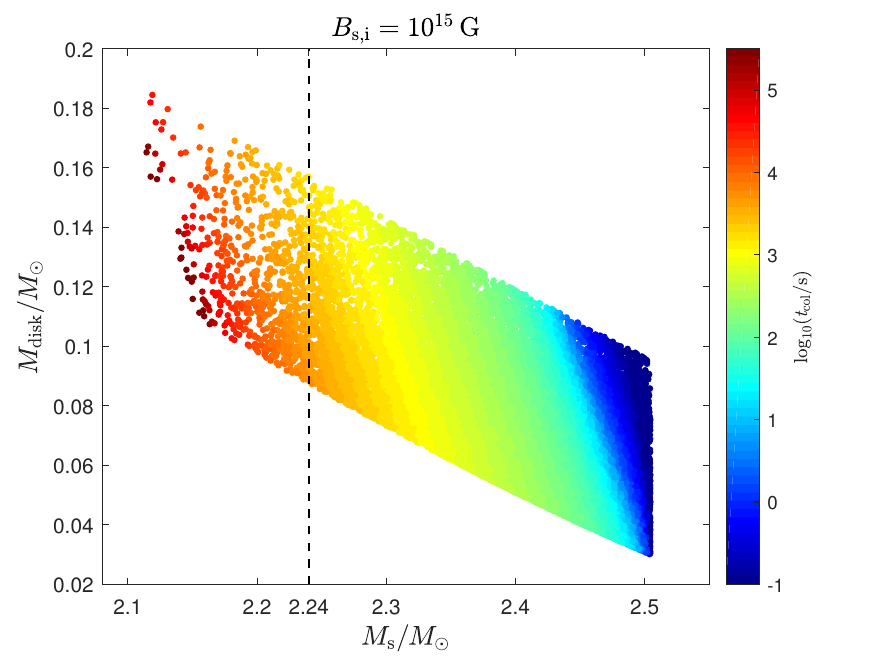}
		\end{minipage}
		\begin{minipage}{0.49\linewidth}
			\includegraphics[height=0.7\textwidth, width=1\textwidth]{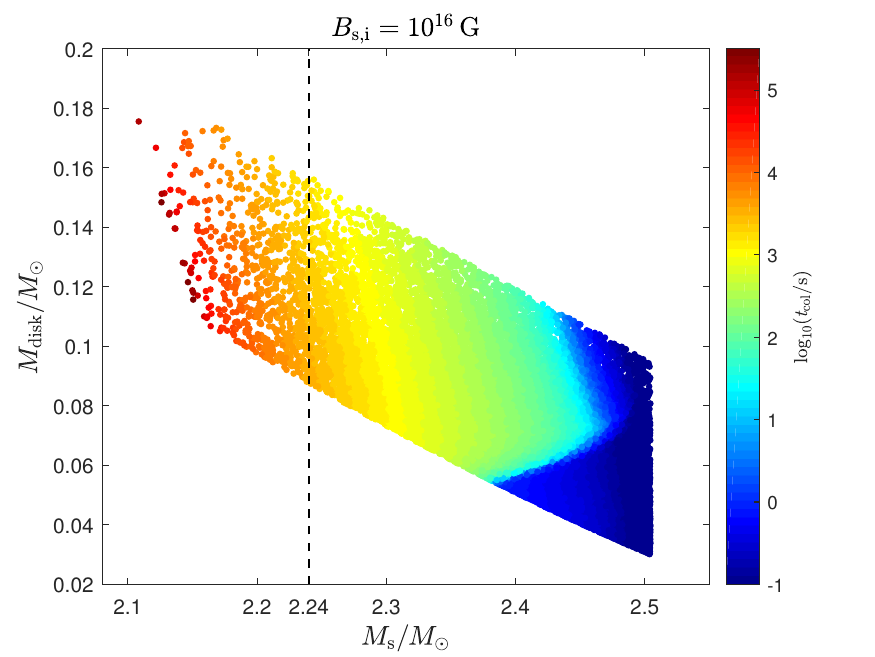}
		\end{minipage}
		
		\begin{minipage}{0.49\linewidth}
			\includegraphics[height=0.7\textwidth, width=1\textwidth]{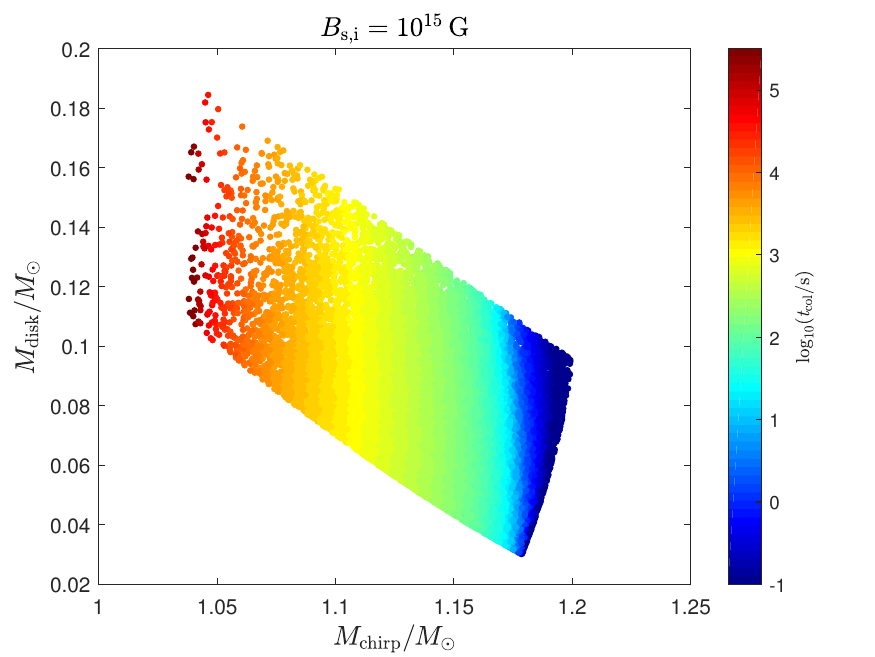}
		\end{minipage}
		\begin{minipage}{0.49\linewidth}
			\includegraphics[height=0.7\textwidth, width=1\textwidth]{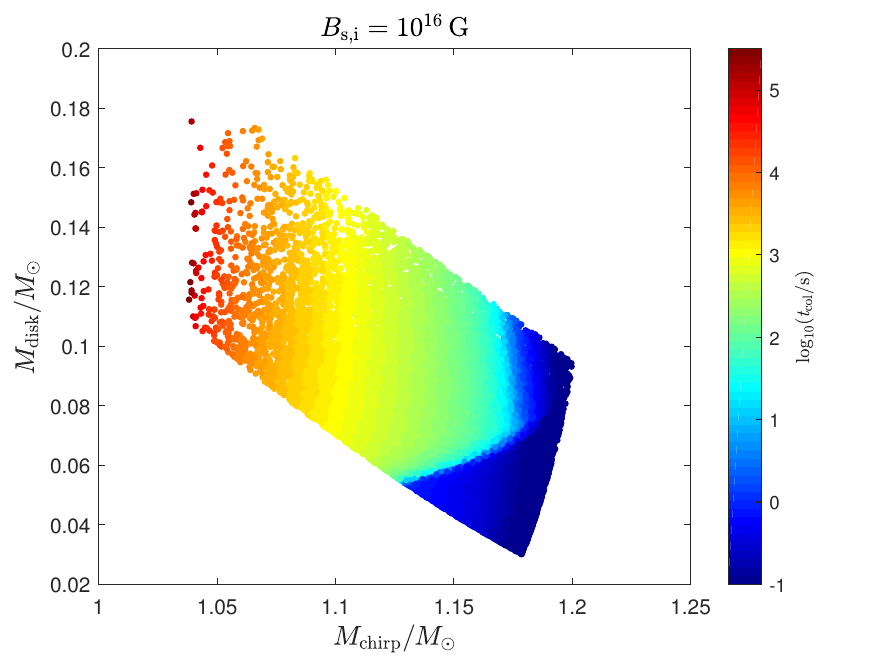}
		\end{minipage}
		\caption{Distributions of the collapse-time (color bar) for the SMNSs from our sample in the $M_{\rm{s}}-M_{\rm{disk}}$
			(upper panels) and $M_{\rm{chirp}}-M_{\rm{disk}}$ (lower panels) diagram. The dashed-line represents the maximum mass of a non-rotating neutron star.}
		\label{fig7:colorbar2}
	\end{figure*}
	
	\begin{figure*}[htb]
		\centering
		\begin{minipage}{0.49\linewidth}
			\includegraphics[height=0.7\textwidth, width=1\textwidth]{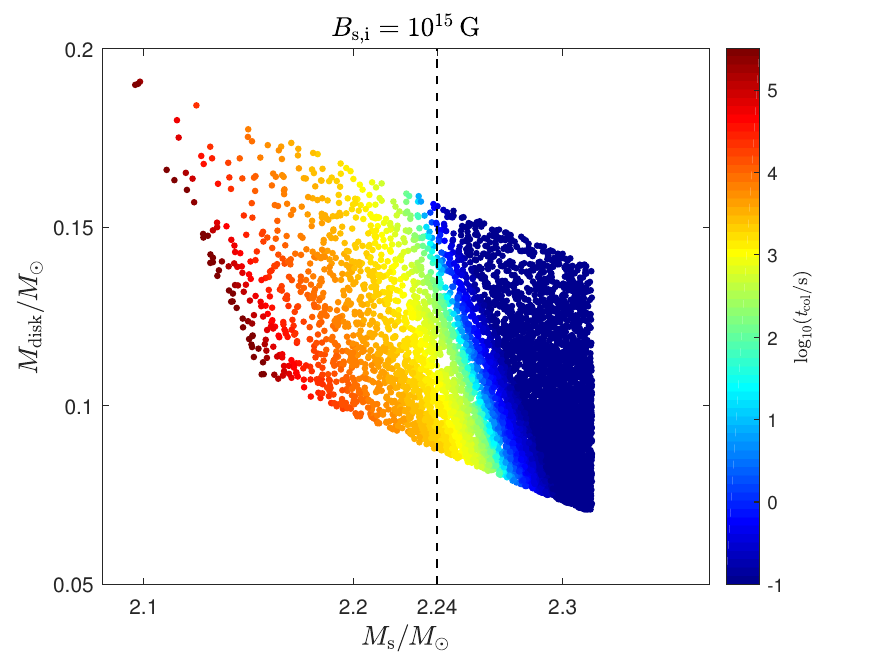}
		\end{minipage}
		\begin{minipage}{0.49\linewidth}
			\includegraphics[height=0.7\textwidth, width=1\textwidth]{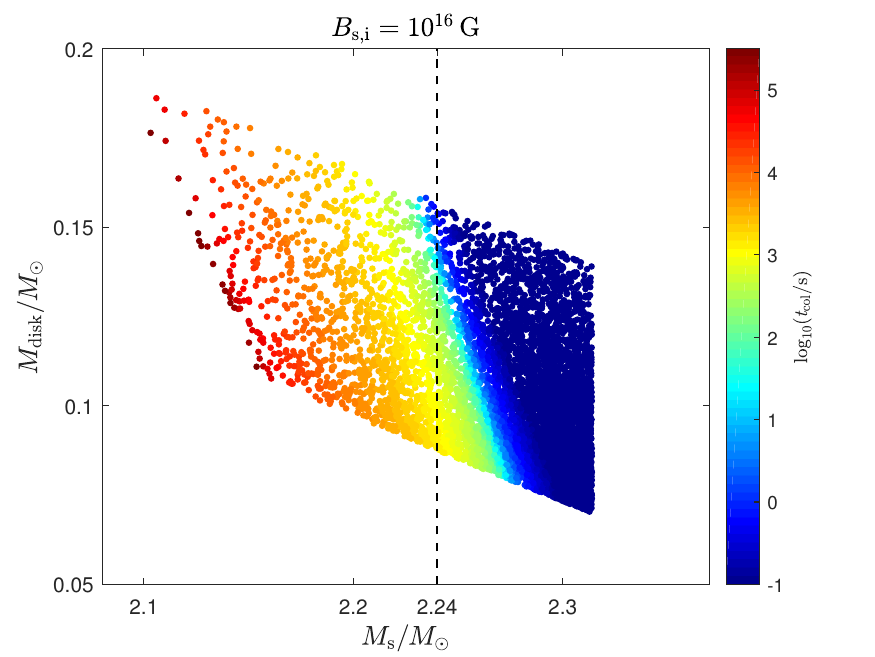}
		\end{minipage}
		
		\caption{The same as Figure \ref{fig7:colorbar2}, but with $P_{\rm{s},i} = 1.5P_{\rm max}$.}
		\label{fig8:colorbar3}
	\end{figure*}

\end{document}